\documentclass[conference]{IEEEtran}
\usepackage{array}
\usepackage{subfigure}
\let\labelindent\relax
\usepackage{enumitem}
\usepackage{fontawesome}
\usepackage[linesnumbered,ruled,vlined]{algorithm2e}
\usepackage{algpseudocode}
\usepackage{multirow}
\usepackage{hhline}
\usepackage{pifont}
\usepackage{bbding}
\usepackage{makecell}
\usepackage{url}
\usepackage{graphicx}
\usepackage{amsmath}
\usepackage{amsfonts}
\usepackage{hyperref}
\usepackage{cleveref}
\usepackage{soul}
\usepackage{xcolor}
\usepackage{colortbl}
\usepackage[switch]{lineno}
\usepackage{tcolorbox}
\tcbuselibrary{listings,breakable}

\hypersetup{
  colorlinks   = true,    
  urlcolor     = red,    
  linkcolor    = red,    
  citecolor    = red      
}

\usepackage{color}
\definecolor{green}{rgb}{0, 0.5, 0}
\definecolor{orange}{rgb}{0.8, 0.6, 0.2}
\definecolor{orange2}{rgb}{1.0, 0.6, 0.2}
\definecolor{red}{rgb}{1.0, 0.0, 0.0}
\definecolor{teal}{rgb}{0.0, 0.4, 0.4}
\definecolor{purple}{rgb}{0.65,0,0.65}
\definecolor{saffron}{rgb}{0.95,0.75,0.2}
\definecolor{turquoise}{rgb}{0.0,0.5,0.5}
\definecolor{black}{rgb}{0.0, 0.0, 0.0}
\definecolor{gray}{rgb}{0.5, 0.5, 0.5}

\newcommand{\name}{\textsc{CamLoPA}\xspace}

\ifCLASSINFOpdf
\else
\fi
\begin{document}

%
\title{\name: A Hidden Wireless Camera Localization Framework via Signal Propagation Path Analysis}



%
\author{\IEEEauthorblockN{Xiang Zhang\IEEEauthorrefmark{1},
Jie Zhang\IEEEauthorrefmark{2},
Zehua Ma\IEEEauthorrefmark{1},
Jinyang Huang\IEEEauthorrefmark{3},
Meng Li\IEEEauthorrefmark{3},
Huan Yan\IEEEauthorrefmark{4},
Peng Zhao\IEEEauthorrefmark{3},
Zijian Zhang\IEEEauthorrefmark{5},
Qing Guo\IEEEauthorrefmark{6}\\,
Tianwei Zhang\IEEEauthorrefmark{2},
Bin Liu\IEEEauthorrefmark{1},
Nenghai Yu\IEEEauthorrefmark{1}}
\IEEEauthorblockA{\IEEEauthorrefmark{1}University of Science and Technology of China \IEEEauthorrefmark{2}Nanyang Technological University \IEEEauthorrefmark{3}Hefei University of Technology}
\IEEEauthorblockA{\IEEEauthorrefmark{4}Guizhou Normal University \IEEEauthorrefmark{5}Beijing Institute of Technology \IEEEauthorrefmark{6}CFAR and IHPC, A*STAR, Singapore}}



\maketitle

\begin{abstract}
Hidden wireless cameras pose significant privacy threats, necessitating effective detection and localization methods. However, existing solutions often require spacious activity areas, expensive specialized devices, or pre-collected training data, limiting their practical deployment. To address these limitations, we introduce \name, a training-free wireless camera detection and localization framework that operates with minimal activity space constraints using low-cost commercial-off-the-shelf (COTS) devices.
\name can achieve detection and localization in just 45 seconds of user activities  with a Raspberry Pi board.
%
During this short period, it analyzes the causal relationship between the wireless traffic and user movement to detect the presence of a snooping camera.
Upon detection, \name employs a novel azimuth location model based on wireless signal propagation path analysis. Specifically, this model leverages the time ratio of user paths crossing the First Fresnel Zone (FFZ) to determine the azimuth angle of the camera. Then \name refines the localization by identifying the camera's quadrant.
%
We evaluate \name across various devices and environments, demonstrating that it achieves 95.37\% snooping camera detection accuracy and an average localization error of 17.46$^o$, under the significantly reduced activity space requirements. Our demo are available at \url{https://www.youtube.com/watch?v=GKam04FzeM4}.

\end{abstract}


%

\section{Introduction}
In recent years, the proliferation of wireless camera devices for home and public security has grown significantly due to their convenience and flexibility in deployment. A study by Market Research Future in 2024~\cite{ankit2024} projected the global wireless video surveillance and monitoring market to grow at a compound annual growth rate of 16.8\% from 2022 to 2030. However, the rapid adoption of wireless cameras has also raised substantial privacy concerns related to unauthorized video recording and dissemination~\cite{ye2013wireless, mare2020smart, David2023}. Users increasingly find themselves being illegally recorded by hidden cameras in various locations, from hotel rooms to short-term rentals. For instance, a 2019 survey~\cite{Jim2019} revealed that 58\% of 2,023 Airbnb guests were concerned about the possibility of hidden cameras, with 11\% reporting actual discoveries of such devices.
In response to these privacy threats, various jurisdictions have proposed and enacted legislation. For example, Delaware's privacy laws now strictly prohibit the use of hidden cameras in private settings without the consent of the individuals being recorded, with violations leading to severe penalties including jail time and fines~\cite{delaware}. These legal measures underscore the urgency of developing effective methods for detecting and localizing hidden wireless cameras.
%

\begin{table}[t]
\centering
\caption{Qualitative comparison with existing approaches.}
\begin{tabular}{c<{\centering}|p{1.3cm}<{\centering}|p{0.8cm}<{\centering}|p{1.4cm}<{\centering}|p{1.1cm}<{\centering}} \hline
 Method & Low-Cost Commercial Devices & Low User Efforts & No Data Pre-Collection and Training & Low Empty Requirement \\ \cline{1-5}
LAPD~\cite{sami2021lapd} & \textcolor{red}{\ding{55}} & \textcolor{red}{\ding{55}} & \textcolor{green}{\checkmark} & \textcolor{green}{\checkmark} \\ \cline{1-5}
HeatDeCam~\cite{yu2022heatdecam} & \textcolor{red}{\ding{55}} & \textcolor{green}{\checkmark} & \textcolor{red}{\ding{55}} & \textcolor{green}{\checkmark} \\ \cline{1-5}
Lumos~\cite{sharma2022lumos} & \textcolor{green}{\checkmark} & \textcolor{red}{\ding{55}} & \textcolor{red}{\ding{55}} & \textcolor{red}{\ding{55}} \\ \cline{1-5}
SNOOPDOG~\cite{singh2021always} & \textcolor{green}{\checkmark} & \textcolor{red}{\ding{55}} & \textcolor{green}{\checkmark} & \textcolor{red}{\ding{55}} \\ \cline{1-5}
MotionCompass~\cite{he2021motioncompass} & \textcolor{green}{\checkmark} & \textcolor{green}{\checkmark} & \textcolor{green}{\checkmark} & \textcolor{red}{\ding{55}} \\ \cline{1-5}
SCamF~\cite{heo2022there} & \textcolor{green}{\checkmark} & \textcolor{red}{\ding{55}} & \textcolor{green}{\checkmark} & \textcolor{red}{\ding{55}} \\ \cline{1-5}
LocCams~\cite{gu2024loccams} & \textcolor{green}{\checkmark} & \textcolor{green}{\checkmark} & \textcolor{red}{\ding{55}} & \textcolor{green}{\checkmark} \\ \cline{1-5}
\name & \textcolor{green}{\checkmark} & \textcolor{green}{\checkmark} & \textcolor{green}{\checkmark} & \textcolor{green}{\checkmark} \\ \hline
\end{tabular}
\label{tab:comint}
\vspace{-0.15in}
\end{table}

Consequently, the problem of wireless camera detection and localization has attracted considerable research attention~\cite{cheng2018dewicam, wu2019you}. However, existing solutions often face significant limitations that hinder their practical deployment. 
Many approaches can detect wireless cameras but cannot locate them~\cite{wu2019you, ji2018user, cheng2019detecting, salman2022csi, dao2021deepdespy}.
Those capable of localization often impose complex requirements. Specifically, methods relying on lens reflection~\cite{sami2021lapd, jakobispy2023, llchidden2023} or electromagnetic/thermal emissions~\cite{liu2023camradar, yu2022heatdecam, zuniga2022see} are typically cumbersome, requiring user expertise and examination of every corner of the room, making them difficult to use. Moreover, electromagnetic/thermal-based methods often necessitate costly specialized equipment.
To address these shortcomings, recent research has focused on analyzing the WiFi traffic and physical layer information to locate wireless cameras. These methods usually require users to move along the edges of the room~\cite{sharma2022lumos, ma2023lenser, heo2022there} or perform perturbations at different positions and orientations~\cite{singh2021always, he2021motioncompass}. The camera's location is determined by assessing the RSSI (Received Signal Strength Indicator) and traffic variations of target devices. These approaches typically necessitate the room to be nearly empty to allow user movement to different locations, which is not feasible in real-world scenarios. They are also time-consuming, requiring 10-30 minutes for camera localization and constant user movement or position adjustments. In a recent work~\cite{gu2024loccams}, differences in WiFi Channel State Information (CSI) under Line-of-Sight (LOS) and None-Line-of-Sight (NLOS) conditions are utilized for the coarse localization of wireless cameras. This approach requires minimal user effort but its localization resolution is limited to 90$^o$, still taking a lot of time to search for devices. Additionally, it requires pre-collected training data, and the deep learning model used has poor robustness against changes in the environment and devices.

\begin{figure}[t]
  \centering
  \includegraphics[width=0.85\linewidth]{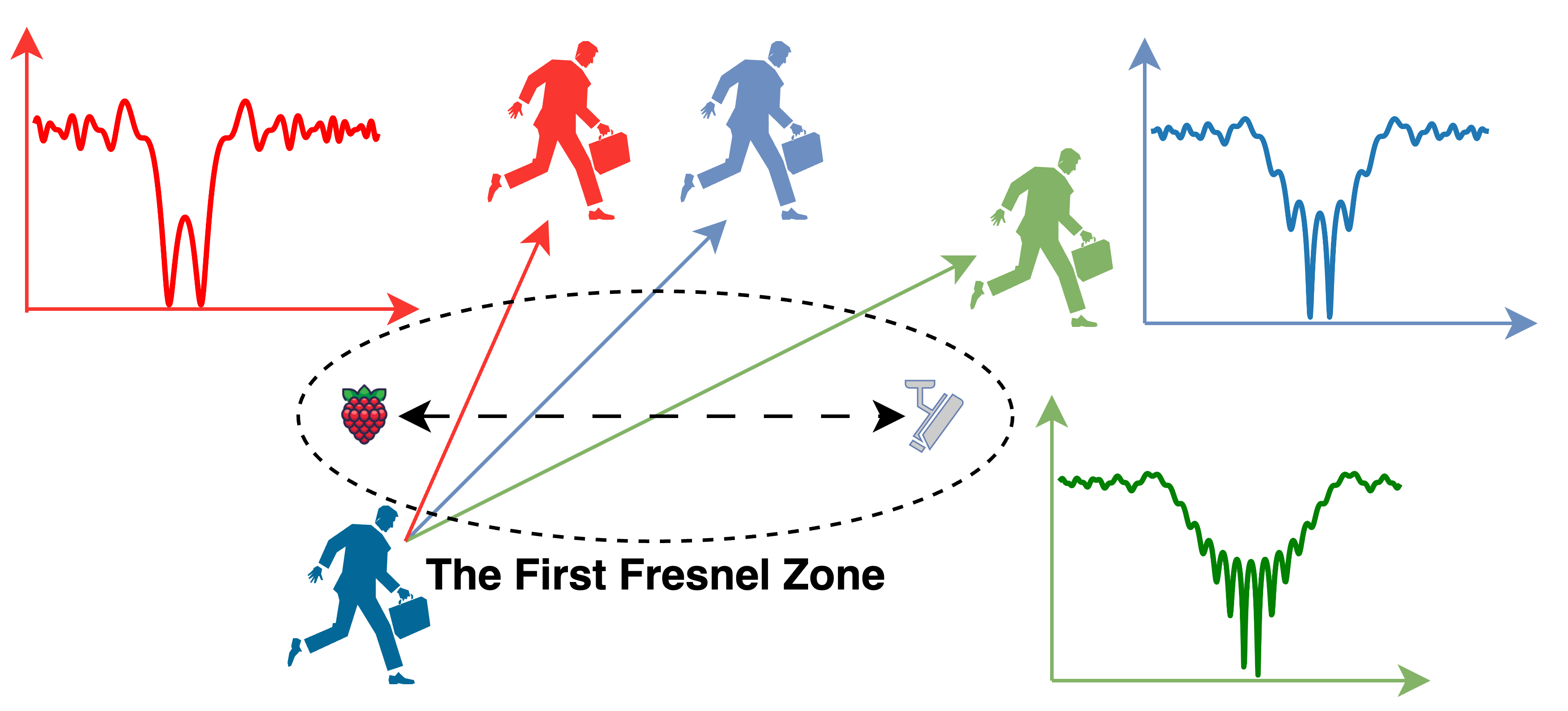}
  \caption{Different wireless signal path losses when crossing the First Fresnel Zone (FFZ) with different path lengthes.}
  \label{fig:infig}
  \vspace{-0.19in}
\end{figure}

In this paper, we introduce \name, a fast and robust wireless camera detection and localization framework using low-cost commercial-off-the-shelf (COTS) devices. As shown in Table~\ref{tab:comint}, \name requires less activity space and user effort compared to previous studies.
\ul{Our framework is inspired by the relationship between obstructions in the propagation path of wireless signals and the resulting signal attenuation.} Specifically, when a large obstacle is located within the First Fresnel Zone (FFZ) between a WiFi transmitter and receiver, the transmitted signal will experience significant attenuation due to diffraction, as defined by Huygen's principle~\cite{baker2003mathematical} and Fresnel-Kirchhoff diffraction parameters~\cite{goldsmith2005wireless}. As illustrated in Figure~\ref{fig:infig}, when a person crosses the FFZ, there is a drastic change in the wireless signal path loss, and the duration of this significant variation is related to the length of the path traversed through the FFZ. Since the FFZ forms an ellipse with the two devices as its foci, given a fixed distance between the two devices, the length of the path through the FFZ can be mapped to the angle of the walk relative to the LOS path (\textbf{azimuth}). \name utilizes this relationship to achieve azimuth angle localization of the wireless cameras.

The technical crux of \name is to address the over-complexity and lack of robustness issues in previous approaches. However, there are two significant challenges:

\noindent\textbf{\textit{1) Relationship Mapping Under Unknown User Speed:}} By analyzing the durations of significant wireless signal fluctuations, we can determine the time it takes for a user to traverse the FFZ. To ascertain the path length through the FFZ, we also need to know the user's speed (The issue of constant user speed is discussed in Section~\ref{sec:diss}.). In real-world scenarios, considering cost and complexity, users typically do not have specialized equipment to measure walking speed or have robots to substitute for user to move. Thus, the user's speed remains unknown, and we cannot determine the path length.
\begin{tcolorbox}[colback=gray!25!white, size=title, breakable, boxsep=1mm, colframe=white, before={\vskip1mm}, after={\vskip0mm}]
\textbf{Q1:} How can we establish a mapping relationship between the traversal time and the azimuth angle of the hidden wireless camera without knowing the user's speed?
\end{tcolorbox}

\noindent\textbf{\textit{2) Errors Control Under Variable Distance and Body Size:}} In practical scenarios, the distance between the hidden wireless camera and the \name device is also unknown, and the user's body size is variable. The user's body size significantly affects the duration of signal variations, as the signal is impacted from the moment the user enters the edge of the FFZ until he/she completely exits from it. Pre-defining these two values
can introduce substantial errors in the aforementioned mapping relationship. 
\begin{tcolorbox}[colback=gray!25!white, size=title, breakable, boxsep=1mm, colframe=white, before={\vskip1mm}, after={\vskip0mm}]
\textbf{Q2:} How can we minimize the impacts of biased parameters and keep the errors within an acceptable range?
\end{tcolorbox}

To overcome the above challenges, we propose a scheme called the \textit{\textbf{orthogonal ratio}}. This scheme replaces the need to measure the distance of a single path through the FFZ with \ul{\textit{the time ratio of two orthogonal paths crossing the FFZ}} to establish a mapping relationship with the azimuth angle. Specifically, we set two orthogonal walking paths that both pass through the \name device, which is typically easy to achieve in real-world environments. We then calculate the time taken for each path to traverse the FFZ. Since the path length is the product of the time and speed, using the time ratio of the two paths eliminates the influence of the speed. Next, we develop a mapping model between the orthogonal ratio and the angle  between the first path and LOS (\textbf{azimuth}) by WiFi propagation path analysis. By obtaining the orthogonal ratio in real environments, the azimuth angle of the wireless camera can be derived from the model. Besides, the orthogonal ratio remarkably reduces the impact of biased parameters such as variable distances and body sizes due to the division operation.

\name operates in three stages and requires only \textbf{45} seconds of user movement to detect and locate a hidden wireless camera. In the first stage (\textbf{0}-\textbf{15}s), the system analyzes the relationship between the data stream uploaded by the camera and user activity for snooping camera detection.
The encoding method of the video stream causes an increase in data volume when there is movement within the monitored area. Therefore, \name first prompts the user to leave the room and collects traffic data of 15 seconds. By examining the causal relationship between the user's exit and the data stream, the system identifies whether a wireless camera is monitoring the current area. 
In the next stage (\textbf{15}-\textbf{35}s), the user walks along two orthogonal paths that both pass through the \name equipment. The system calculates the orthogonal ratio of these two paths and determines the azimuth of the wireless camera using the azimuth model. This model only provides an angle within the range of 0-90° (e.g., for 45° and 135°, \name reports 45° for both cases). To address this, we further design a scheme to determine the quadrant in which the camera is located. 
In the final stage (\textbf{35}-\textbf{45}s), the system prompts the user to walk along a path that coincides with the first path but does not traverse the entire FFZ. By analyzing whether the user's initial position blocks the LOS, the quadrant determination scheme identifies the quadrant in which the wireless camera is located, achieving the final localization. We implement a prototype of \name on a Raspberry Pi device, which users can connect to using SSH tools on their smartphone to receive system prompts and display the results. 

In summary, we make the following key contributions:

\begin{itemize}[leftmargin=*]
	\item We propose \name, the first hidden wireless camera detection and localization framework based on the diffraction phenomenon during wireless signal propagation. This scheme is implemented using low-cost COTS devices. It has small activity space requirements, and does not require model training.
	\item We introduce a wireless device azimuth localization model and a quadrant determination method based on wireless signal propagation path analysis. The model is designed on the principle that diffraction causes significant attenuation of wireless signals. By combining the model with the quadrant determination method, we can achieve fast and training-free device localization.
	\item We evaluate \name across various devices and environments. Experiment results show that \name achieves the detection accuracy of 95.27\% and average localization error of 17.46° for snooping wireless cameras.
\end{itemize}

\section{Background}

We first provide an overview of the state-of-the-art approaches for detecting and locating hidden cameras. Next, we introduce Channel State Information (CSI), which can be used to characterize signal attenuation.

\subsection{Detecting and Locating Hidden Wireless Cameras}

Current wireless hidden camera detection schemes generally rely on information leaked from wireless channels or other side channels when the camera is in operation. For example, wireless communication can unintentionally leak information in certain out-of-band channels, which has recently been used to detect the presence of wireless devices, however, Sathyamoorthy et al.~\cite{sathyamoorthy2014wireless} and Valeroset al.~\cite{valeros2017spy} propose that the received power threshold needs to be carefully set to avoid false alarms or detection failures. LAPD~\cite{sami2021lapd}, CamRadar~\cite{liu2023camradar} and Heatdecam~\cite{yu2022heatdecam} are based on thermal/electromagnetic emissions and lens reflection side channels when the camera is operating. These methods usually utilize relatively expensive specialized sensors to capture such information for detection. While they can locate devices within the LOS, they require the detection equipment to be close to the hidden camera to capture subtle changes in these side-channel signals, making them challenging for ordinary users and ineffective for detecting devices in unreachable areas. Some methods leverages WiFi packet sniffing to detect wireless cameras, as these cameras transmit data packets when in operation. Dewicam~\cite{cheng2018dewicam}, Cheng et al.~\cite{cheng2019detecting}, Liu et al.~\cite{liu2018detecting} and Miettinen et al.~\cite{miettinen2017iot} achieved detection by learning the traffic characteristics of wireless cameras, but machine learning-based approaches often suffered from robustness issues due to their reliance on large datasets. SNOOPDOG~\cite{singh2021always} and ScamF~\cite{heo2022there} relied on the causal relationship between wireless camera traffic and human activity, where significant movement within the monitored area led to increased encoded data. This semantic-rich relationship helped determine whether an area was under surveillance. Motioncompass~\cite{he2021motioncompass} and LocCams~\cite{gu2024loccams} involved side-channel information collected from the internet, such as the Organizationally Unique Identifier (OUI) in the Media Access Control (MAC) address of WiFi packets, which often included information about the device's manufacturer. This information can be used to identify the type of device.

The localization of wireless hidden cameras also relies on the leakage of side-channel information, but not all side-channel information can be used simultaneously for detection and localization. Schemes based on thermal/electromagnetic emissions~\cite{liu2023camradar,yu2022heatdecam} and lens reflection~\cite{sami2021lapd} side channels can detect and localize cameras by directly capturing the regions with abnormal signals. However, their limitations for localization are similar to those for detection: they are difficult to use and require proximity to the hidden camera~\cite{gu2024loccams}. Detection schemes based on traffic analysis require additional effort to achieve localization. For example, they may rely on changes in RSSI strength or data flow as the user carrying the detection device moves around to infer the camera's location~\cite{singh2021always,sharma2022lumos,heo2022there}. These schemes necessitate that the room be almost empty, which may not be feasible in everyday settings with furniture, as the user's movement space is limited and they might not be able to get close to the hidden wireless camera. Recently, LocCams~\cite{gu2024loccams} has been proposed that uses CSI to determine whether the user is blocking the LOS path between the positioning equipment and the wireless camera, thus roughly estimating the hidden camera's location. However, this method has a localization resolution of only 90 degrees, and its deep learning-based approach suffers from poor robustness without the support of a large-scale dataset.

\subsection{Channel State Information (CSI)}
\label{subsec:csii}

WiFi CSI describes various effects that a WiFi signal undergoes during propagation, including multipath effects, attenuation, phase shift, and more. This process of influence can be represented as follows~\cite{huang2023phyfinatt,zhang2023wital}:
	\begin{equation}
	\label{equ:CH}
	Y = H \cdot X+ N,
	\end{equation}
where $Y$ and $X$ are the received and transmitted signals, respectively. $N$ is the additive white Gaussian noise, and $H$ is a complex matrix representing CSI. And this complex matrix can be expressed as follows:
	\begin{equation}
	\label{equ:fH}
	H(f) = |H(f)| e^{j\theta(f)},
	\end{equation}
where $H(f)$ is the channel response at frequency $f$, $|H(f)|$ is the magnitude of the CSI, representing the variation in signal strength, and $\theta(f)$ is the phase shift of the CSI, representing the variation in signal phase. The magnitude of the CSI can be used to characterize signal attenuation. The received CSI is a superposition of signals of all the propagation paths, and its Channel Frequency Response (CFR) can be represented as~\cite{zhang2024wiopen}:
\begin{equation}
\label{equ:CSI}
H(f,t)=\sum_{m\in \Phi }^{} a_{m}(f,t) e^{-j2\pi \frac{d_{m}(t) }{\lambda } },
\end{equation}
where $f$ and $t$ represent center frequency and time stamp, respectively, and $m$ is the multi-path component. $a_{m}(f,t)$ and $d_{m}(t)$ denote the complex attenuation and propagation length of the $m$th multi-path component, respectively. $\Phi$ denotes the set of multi-path components and $\lambda $ is the signal wavelength. When there are changes in only one path, the CSI can be used to approximate the attenuation occurring on that path. Specifically, paths with no changes and those with changes can be categorized as static and dynamic paths as follows~\cite{gu2023wife}:
\begin{equation}
	\begin{aligned}
		\label{equ:DNS}
		H(f,t) &= H_{s} (f,t)+H_{d} (f,t)\\
		&=\sum_{m_{s}\in \Phi_{s} }^{} a_{m_{s}}(f,t) e^{-j2\pi \frac{d_{m_{s}}(t) }{\lambda } }\\
		& + \sum_{m_{d}\in \Phi_{d} }^{} a_{m_{d}}(f,t) e^{-j2\pi \frac{d_{m_{d}}(t) }{\lambda } },
	\end{aligned}
\end{equation}
where $H_{s}(f,t)$ and $H_{d}(f,t)$ denote the static and dynamic components, respectively. $\Phi_{s}$ represents the set of static paths, e.g., reflected off the walls and furniture and static body parts, while $\Phi_{d}$ denotes the set of dynamic paths, e.g., reflected off the moving human. When there is only one person moving in the room, CSI can be used to characterize the signal attenuation and multipath effects caused by this person's movement.

\begin{figure}[t]
  \centering
  \includegraphics[width=0.95\linewidth]{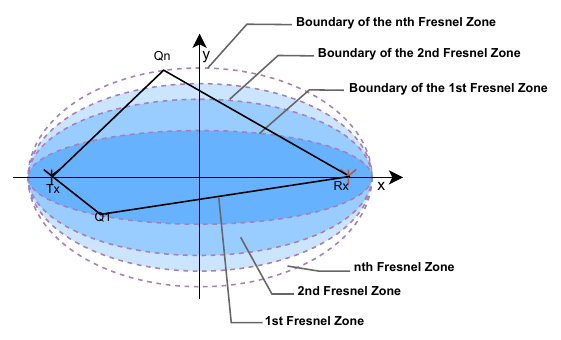}
  \caption{Illustration of Fresnel Zone.}
  \label{fig:fzone}
  \vspace{-0.19in}
\end{figure}

\begin{figure*}[t]
  \centering
  \includegraphics[width=1\linewidth]{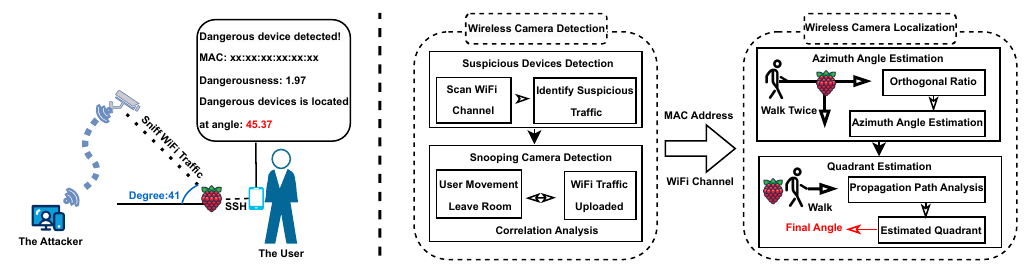}
  \caption{Overview of \name. \name is implemented using a low-cost Raspberry Pi, which can connect via SSH to the user's phone for prompts and notifications. The operation of \name is divided into two phases: wireless camera detection and localization. The detection stage determines whether a wireless camera is monitoring the current area, while the localization stage precisely locates the identified camera.}
  \label{fig:camlopa}
  \vspace{-0.15in}
\end{figure*}

Next, we briefly explain the Fresnel zone model, which is widely used to analyze the diffraction and reflection effects of wireless and light signals along their propagation path. This model helps in understanding how signal strength varies with distance and obstacles. The Fresnel zones can be described as a series of concentric ellipses with the wireless signal transmitter and receiver as the focal points~\cite{rappaport2024wireless} (see Figure~\ref{fig:fzone}):
\begin{equation} 
	\label{equ:fz}
	|TxQ_{n}|+|Q_{n}Rx|-|TxRx|=n\lambda/2,
\end{equation}
where $Q_n$ is a point at the boundary of the $n$th Fresnel zone, and $Tx$ and $Rx$ represent the transmitter and receiver, respectively. Since the phase difference of waves within the First Fresnel Zone (FFZ) is relatively small, most of the energy is concentrated in this region. In wireless communication and wave propagation, the energy within the FFZ typically accounts for about 60\% to 70\% of the total transmitted energy. Obstacles outside the FFZ primarily cause signal reflection~\cite{zhang2019towards,yao2024wiprofile,wang2024understanding}. The attenuation due to reflection is minimal, and the total signal energy affected by obstacles outside the FFZ is relatively small. As a result, when obstacles moves in the outside of the FFZ, the total received signal energy does not change significantly. Instead, the movement mainly causes multipath effects, leading to phase changes in the CSI. Conversely, obstacles within the FFZ mainly cause diffraction~\cite{goldsmith2005wireless,rappaport2024wireless}. The attenuation due to diffraction is substantial, and since a significant amount of signal energy is transmitted within the FFZ, the received signal experiences substantial attenuation, which can be clearly characterized by the magnitude of the CSI.


In practical systems, we can use open-source tools such as csitool~\cite{halperin2011tool}, picosense~\cite{lireshaping}, and nexmon\_csi~\cite{gringoli2019free,nexmon:project} to obtain CSI from various network cards, including Intel 5300, AX210/AX200, and bcm43455c0 (Raspberry Pi B3+/B4). The actual size of the extracted CSI matrix depends on the number of antennas and subcarriers~\cite{ma2019wifi,gu2022wigrunt}, and the obtained CSI is a 4-dimensional tensor $H\in \mathbb{C^{N\times M\times K\times T}}$, and $\mathbb{M}$, $\mathbb{K}$, and $\mathbb{T}$ represent the number of receive antennas, transmit antennas, subcarriers, and packets, respectively.

\section{Overview}

\subsection{Threat Model}

Our work focuses on an attacker who places a hidden wireless camera in a room to monitor users. This aligns with current state-of-the-art methods~\cite{wampler2015information,nassi2019drones,gu2024loccams,heo2022there,sharma2022lumos} for detecting and locating any hidden camera. This is also supported by several cases~\cite{Fussell2019,Jeong2019,10587029} where attackers were live-streaming users in private spaces, a convenient and practical solution to gather users' private information. Therefore, we focus on the scenarios that the attacker uses COTS wireless cameras for privacy invasion. The adversary covertly deploys a camera in the room and communicates with it using encrypted wireless traffic. We focus on WiFi as the communication channel in this paper, as it is the most common method for remote monitoring with commercial and consumer devices. Below we introduce the real-world settings for both the attacker and user.

\noindent\textbf{Attacker:} The attacker could be the host or a previous guest intending to monitor users in the room. 
\begin{itemize}[leftmargin=*]
	\item The attacker can fully control the room before the user checks in, such as changing the environment and installing the hidden wireless camera.
	\item The attacker uses a COTS camera device to spy on users and can control the camera through an app. Similar to previous studies~\cite{ortiz2019devicemien, sivanathan2018classifying, sharma2022lumos, singh2021always, heo2022there}, we assume the attacker does not alter the firmware, network protocols or wireless transmission behaviors of the camera device, as these tasks generally require a high level of expertise.
	\item The attacker has complete control over the WiFi network to which the hidden wireless camera connect. He can configure the WiFi network's wireless channels, encryption methods, and access modes.
\end{itemize}

\noindent\textbf{User:} The user's requirement is to detect and locate the hidden wireless camera within the room.
\begin{itemize}[leftmargin=*]
	\item The user can access the physical space to search and move around. But in a real environment, his movement is limited and obstructed by the furniture, making it difficult to meet the activity space requirements of most previous studies~\cite{heo2022there, sharma2022lumos, he2021motioncompass, singh2021always}.
	\item The user does not have any knowledge of the hidden wireless camera. He is unaware of the WiFi network being used, the channel of the WiFi network, or the camera's location. However, the user has control over the \name device, including its placement and the configuration of its network connection.
	\item The user does not have control over the WiFi network to which the wireless camera are connected. However, he can use existing tools (e.g., tcpdump, Wireshark) to sniff WiFi 802.11 packets broadcast in the air. The user carries no additional measuring tools except for a Raspberry Pi equipped with \name.
\end{itemize}

\subsection{Workflow of \name}
\name requires the user to perform three walks (45 seconds) to detect and locate the hidden wireless camera according to the prompts of \name. It then provides feedback with the estimated azimuth angle of the hidden wireless camera. The overall structure of \name is shown in Figure~\ref{fig:camlopa} and it operates in two phases:

\noindent\textbf{Hidden Wireless Camera Detection.}
\name first scans the surrounding WiFi networks and captures packets on all active 802.11 wireless channels for analysis. If it detects a device that is continuously uploading data, it identifies this device as suspicious and forwards its MAC address and channel index to the snooping camera detection module. The snooping camera detection module will prompt the user to leave the room and sniff packets from this channel for 15 seconds. It then analyzes the upload traffic of the suspicious device according to the MAC address. If the traffic pattern matches the user's departure phase, the detection module will report that the device is monitoring the current area. Next, the module will forward the device's MAC address and channel index to the following localization phase.

\noindent\textbf{Hidden Wireless Camera Localization.}
Upon receiving the MAC address of the snooping wireless camera and the WiFi channel of the connected Access Point (AP), \name prompts the user to walk along two orthogonal paths (see Figure~\ref{fig:amizuth}) cross the \name device, such as a Raspberry Pi board. Specifically, the device sniffs the WiFi packets transmitted from the target MAC on the specified channel over 10 seconds for each path, extracting CSI to calculate the orthogonal ratio and determine the azimuth angle using the proposed azimuth localization model. These paths intersect in a T-shape, with the intersection point being the location of the \name device. After calculating the azimuth angle, \name prompts the user to walk along a path coinciding with the first path but starting in front of the \name device, collecting 10 seconds of CSI. Next, using the quadrant determination model, \name calculates the quadrant in which the target device is located to obtain the final azimuth angle of the hidden wireless camera.

\section{Wireless Camera Detection}

\name detects the presence of snooping wireless cameras in the environment through wireless traffic analysis by: (i) searching for suspicious devices, and (ii) detecting snooping wireless cameras.

\subsection{Searching for Suspicious Devices}

In real-world environments, there are usually many wireless networks and devices connected to WiFi around the user. Analyzing all devices to detect cameras monitoring the area is highly inefficient. Therefore, \name first identifies suspicious devices to narrow down the detection scope. Video stream packets are typically large and stable, and surveillance cameras continuously and frequently upload data. \name starts by scanning the surrounding WiFi networks to detect all APs, even those with Hidden Service Set Identifiers (SSIDs). According to~\cite{meta2024}, we list the Received Signal Strength Indication (RSSI) requirements for different applications. \name excludes APs that do not meet the minimum RSSI requirements for video streaming, namely, below -67 dBm. It then sequentially scans the channels of the remaining APs, sniffing and capturing 802.11 packets for 5 seconds to determine if any devices are continuously uploading data.

\begin{table}[t]
\centering
\caption{Received Signal Strength Indication (RSSI).}
\begin{tabular}{p{0.7cm}|p{1cm}|p{3.8cm}|p{1.3cm}} \hline
 Signal Strength & Conclusion & Describe & Required for \\ \cline{1-4}
 -30 dBm & Amazing & Max achievable signal strength. Not typical or desirable in the real world. & N/A\\ \cline{1-4}
 -67 dBm & 	Very Good & 	Minimum signal strength for applications that require very reliable, timely delivery of data packets. & VoIP, streaming video\\ \cline{1-4}
 -70 dBm & 	Okay & 	Minimum signal strength for reliable packet delivery. & Email, web\\ \cline{1-4}
 -80 dBm & 	Not Good & 	Minimum signal strength for basic connectivity. Packet delivery may be unreliable. & N/A\\ \cline{1-4}
 -90 dBm &  Unusable & 	Approaching or drowning in the noise floor. Any functionality is highly unlikely. & N/A\\ \cline{1-4}
\end{tabular}
\label{tab:rssiapp}
\end{table}

\begin{figure}[t]
  \centering
  \includegraphics[width=0.82\linewidth]{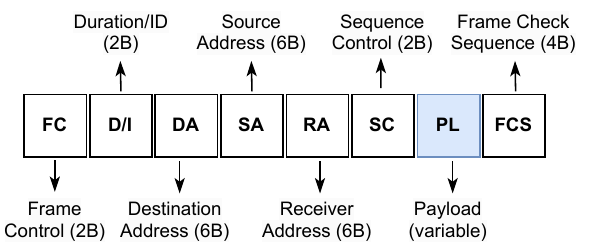}
  \caption{IEEE 802.11 wireless frame.}
  \label{fig:802.11f}
  \vspace{-0.15in}
\end{figure}

For the captured 802.11 packets, \name first classifies them by source MAC address into different end devices. Next, it filters out Management-Type and Control-Type frames, leaving only Data-Type frames for further analysis, as application layer data is encapsulated within Data-Type frames~\cite{li2022packet}. The structure of an 802.11 wireless frame~\cite{gast2005802,9363693} is shown in Figure~\ref{fig:802.11f}. It includes the plaintext header information and the encrypted data payload. The header contains unencrypted information such as addresses, while the payload is typically encrypted using WEP/WPA/WPA2. After protocol filtering, \name aggregates all Data-Type frames corresponding to each device and calculates the average size of the payload portion. Finally, \name determines the presence of any suspicious devices as follows:
\begin{equation}
\label{sdfind}
	\mathbf{S_{mac}} =\begin{cases}
	\mathbf{true}  & \text{ if } \bar{s} _{mac}>T_s \&  l>T_l  \& \mathbf{mac}\ne \mathbf{m_{ap}},\\
	\mathbf{false}  & \text{ else }.
	\end{cases}
\end{equation}
Here, $\mathbf{S_{mac}}$ represents the determination of whether the device with MAC address $\mathbf{mac}$ is suspicious. $\bar{s}_{mac}$, $T_s$, $l$, $\mathbf{m_{ap}}$, and $T_l$ denote the average size of all packet payloads, the size threshold, the count of packets, the MAC address of APs, and the count threshold, respectively. This equation indicates that if a device sends a large number of packets within 5 seconds and the average packet length is long, it is likely uploading a video stream. After identifying suspicious devices, \name forwards their MAC addresses and 802.11 channel index to the snooping camera detection module. This module then sequentially assesses the risk of each device to determine whether they are monitoring the current area.

\subsection{Detecting Snooping Cameras}

Before uploading video streams, cameras typically apply encoding to compress the data and reduce the upload volume. Most video compression standards, such as H.264~\cite{van2008traffic} and H.265~\cite{pan2016fast}, achieve high compression rates through inter-frame prediction. Specifically, standard video compression algorithms use three types of frames to compress video:
1) I (Intra-coded picture) frames: these frames contain complete image information and can be decoded independently of other frames.
2) P (Predicted picture) frames: these frames encode residual information and require information from preceding I frames for decoding.
3) B (Bi-directionally predicted picture) frames: these frames can construct images using changes from preceding I or P frames, subsequent I or P frames, or interpolations between preceding and subsequent I/P frames.
Among these frame types, B frames are the most compressible, followed by P frames, and finally, I frames.

\begin{figure}[t]
  \centering
  \includegraphics[width=0.95\linewidth]{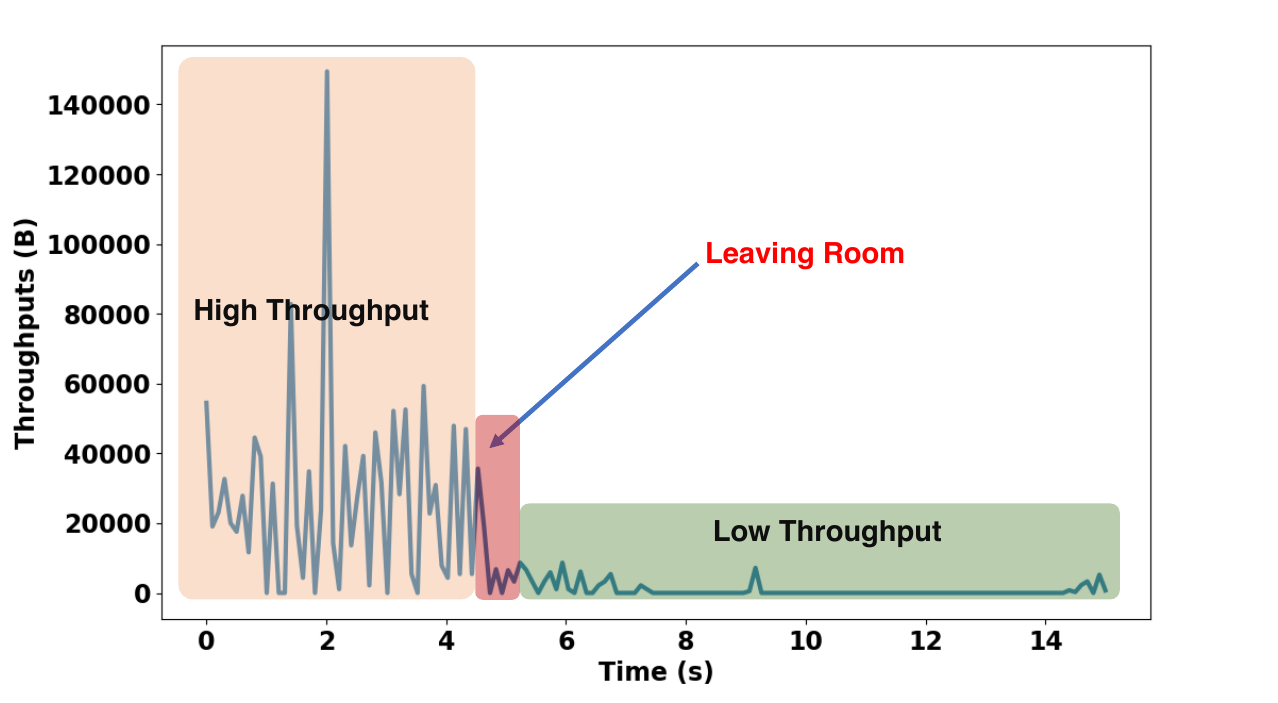}
  \caption{Throughput during the user’s exit from the room.}
  \label{fig:throu}
  \vspace{-0.19in}
\end{figure}

Therefore, when there is any activity in the area monitored by the wireless camera, the camera traffic increases due to the higher number of P and B frames that need to be transmitted~\cite{singh2021always,heo2022there}. Conversely, if the scene transitions to a stationary one, the number of disturbed pixels decreases, reducing the camera traffic. If a person first moves and then remains still within the camera's monitored area, it will result in a unique camera traffic pattern (traffic decreasing) that corresponds to the user's motion. This causal effect can be used to detect whether a hidden wireless camera is snooping on the current area. \name leverages this causal relationship to detect snooping cameras. Specifically, \name prompts the user to leave the room within 15 seconds. It then calculates the data throughput of each suspicious device per second and checks for traffic patterns where the throughput is initially high and then decreases. If such a pattern is detected, the device is identified as a snooping camera, and its risk level is determined based on the ratio of the data throughput in the first half to that in the second half. A sample of the data throughputs during the user's exit from the room is shown in Figure~\ref{fig:throu}.

Upon detecting a snooping camera, \name forwards the camera's MAC address and associated WiFi channel index to the wireless camera localization module. It then initiates the localization process for the detected snooping camera.

\section{Wireless Camera Localization}

\name localizes snooping cameras in two stages: (i) azimuth localization and (ii) quadrant determination.

\subsection{Diffraction Attenuation in Wireless Signal Propagation}
\label{subsec:diffatt}

Diffraction allows radio signals to propagate around the curved surface of the earth, beyond the horizon, and behind obstacles~\cite{rappaport2024wireless}. This phenomenon can be explained using Huygen's principle, which states that all points on a wavefront can be considered as point sources generating secondary wavelets. These secondary wavelets are combined in the direction of propagation to form a new wavefront. Diffraction occurs due to the propagation of these secondary wavelets into shadowed regions. Empirical studies~\cite{zhang2019towards,zhang2018,wang2024understanding} suggest that when an obstacle is within the FFZ, it primarily causes the diffraction of wireless signals. Conversely, when the obstacle is outside the FFZ, it mainly causes the reflection of signals.

\begin{figure}[t]
  \centering
  \includegraphics[width=0.7\linewidth]{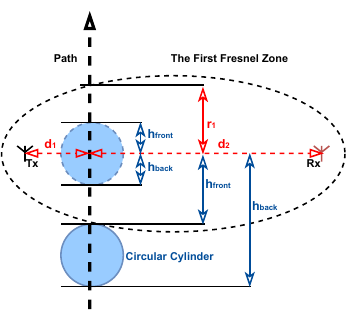}
  \caption{A moving cylinder across the FFZ.}
  \label{fig:fzonecy}
  \vspace{-0.2in}
\end{figure}

In Figure~\ref{fig:fzone}, assuming the height of point $Qn$ from the LOS path is $h$ (see Figure~\ref{fig:fzonecy}), and its projection onto the LOS path has distances $d_1$ and $d_2$ from $Tx$ and $Rx$, respectively, the path difference between the signal propagating through this point and the LOS path $\Delta d$ can be expressed as~\cite{rappaport2024wireless}:
\begin{equation}
	\Delta d \approx \frac{h^2}{2} \frac{d_1+d_2}{d_1d_2}.
\end{equation}
The corresponding phase difference is:
\begin{equation}
\label{equ:phidif}
	\phi = \frac{2\pi d}{\lambda} = \frac{\pi h^2}{\lambda} \frac{d_1+d_2}{d_1d_2}.
\end{equation}
Equation~\ref{equ:phidif} can typically be expressed using the Fresnel-Kirchoff diffraction parameter $v$ as follows:
\begin{equation}
	\phi = \frac{\pi}{2}v^2.
\end{equation}
The Fresnel-Kirchoff diffraction parameter $v$ can be represented as:
\begin{equation}
	v = h\sqrt{\frac{2(d_1+d_2)}{\lambda d_1d_2}}.
\end{equation}
The Fresnel-Kirchoff diffraction parameter originates from the combination of the Fresnel approximation and Kirchhoff’s diffraction theory. This parameter is used to describe the diffraction effect that occurs when a wave encounters an obstacle or aperture. The magnitude of $v$ is related to the significance of the diffraction effect. A smaller $v$ indicates a smaller obstacle size or greater distance, resulting in a less significant diffraction effect. Conversely, a larger $v$ indicates a more pronounced diffraction effect, where the wave experiences noticeable diffraction when encountering an obstacle and continues to propagate around it.
The radius (The perpendicular distance from $Q_n$ to the LOS path.) of the FFZ can be expressed as~\cite{rappaport2024wireless}:
\begin{equation}
	r_1 = \sqrt{\frac{\lambda d_1d_2}{d_1+d_2}}.
\end{equation}
Thus, the Fresnel-Kirchoff diffraction parameter can be represented as:
\begin{equation}
	v=h\sqrt{\frac{2(d_1+d_2)}{\lambda d_1d_2}}=h\frac{\sqrt{2}}{r_1}.
\end{equation}

In wireless communication systems, only a portion of the signal's energy can diffract around an obstacle, allowing only part of the blocked energy to reach the receiver. Therefore, when an obstacle obstructs part of the Fresnel zone, the received energy is the vector sum of the contributions from all the unobstructed portions of the Fresnel zone. If an infinitely long object is positioned at a distance $h$ from the LOS path, the ratio of the electric field strength $E_d$ affected by diffraction to the unobstructed electric field strength $E_o$ is given by:
\begin{equation}
	\frac{E_d}{E_o} = F(v) = \frac{1+j}{2}\int_{v}^{\infty}exp(\frac{-j\pi t^2}{2})dt,
\end{equation}
where $F(v)$ is the complex Fresnel integral.

In practical scenarios, a human body can be approximated as a cylinder to analyze the signal attenuation caused by diffraction along the propagation path. As shown in Figure~\ref{fig:fzonecy}, both ends of the cylinder induce diffraction effects, where  $h_{\text{front}}$  and  $h_{\text{back}}$  represent the distances from the front and back edges of the cylinder to the LOS path, respectively. The signal attenuation caused by diffraction at the front and back edges can be expressed as:
\begin{equation}
	F(v_{front}) = \frac{1+j}{2}\int_{v_{front}}^{\infty}exp(\frac{-j\pi t^2}{2})dt ,
\end{equation}
\begin{equation}
	F(v_{back}) = \frac{1+j}{2}\int_{-\infty}^{v_{back}}exp(\frac{-j\pi t^2}{2})dt .
\end{equation}
The diffraction gain due to the presence of a cylinder is given by:
\begin{equation}
	G_d(dB)=20log|F(v_{front})+F(v_{back})|.
\end{equation}

\begin{figure}[t]
  \centering
  \includegraphics[width=0.92\linewidth]{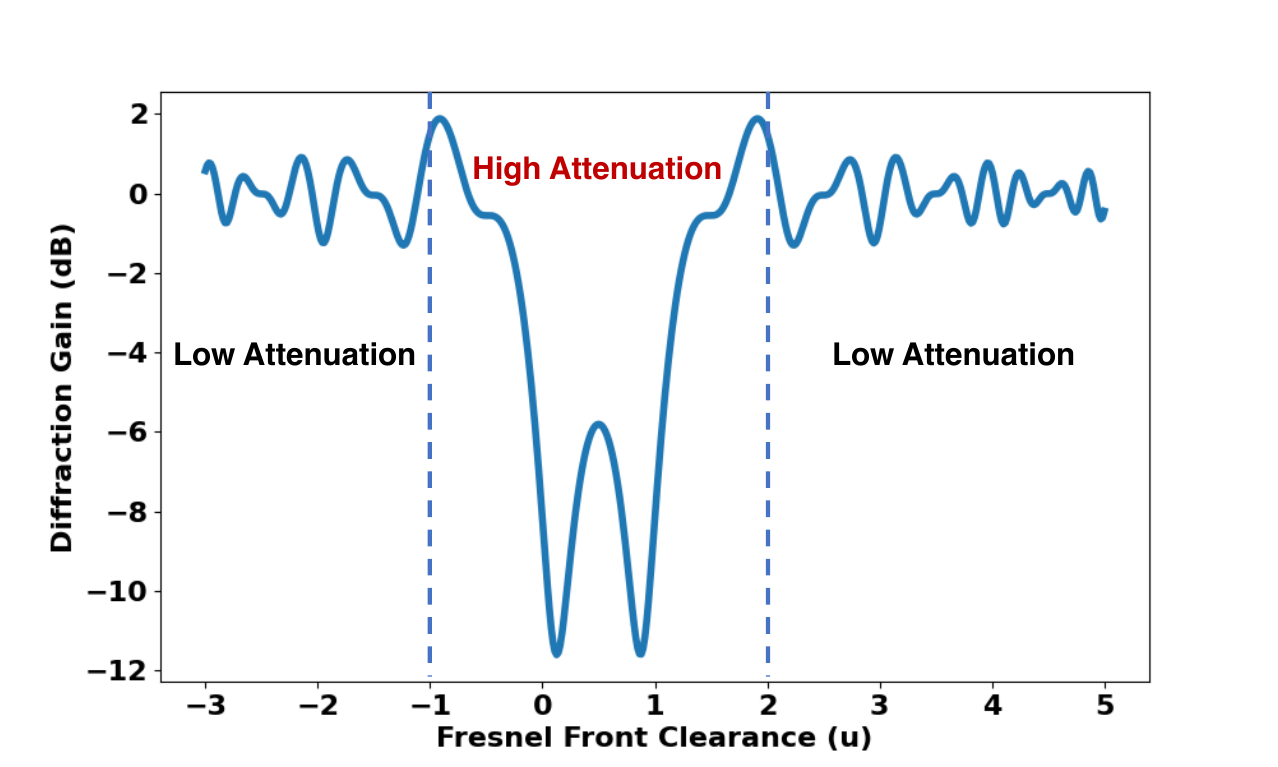}
  \caption{Diffraction gain variation corresponding to Figure~\ref{fig:fzonecy}.}
  \label{fig:digain}
  \vspace{-0.2in}
\end{figure}

To intuitively demonstrate the diffraction attenuation caused by obstruction, we use the example of a cylinder with a radius equal to the FFZ radius. To simplify the setup, we assume the cylinder crosses the FFZ vertically (as shown in Figure~\ref{fig:fzonecy}) and introduce Fresnel clearance $u$~\cite{zhang2018} to indicate the percentage of crossing:
\begin{equation}
	u=\frac{h}{r_1},
\end{equation}
\begin{equation}
	v=h\sqrt{\frac{2(d_1+d_2)}{\lambda d_1d_2}}=h\frac{\sqrt{2}}{r_1} = \sqrt{2}u.
\end{equation}
The diffraction gain during the cylinder's traversal of the FFZ is shown in Figure~\ref{fig:digain}. It is obvious that the cylinder causes significant signal attenuation due to diffraction from the moment it touches the FFZ ($u_{front}=-1$) until it completely exits the FFZ ($u_{front}=2$).

\subsection{Azimuth Localization}

Section~\ref{subsec:diffatt} highlights that the period of significant wireless signal attenuation can be used to determine the time taken for an obstacle (the user) to cross the first Fresnel zone (FFZ). Below, we list several key points:
\begin{itemize}
	\item The location of the \name device is known.
	\item As discussed in Section~\ref{subsec:csii}, Channel State Information (CSI) can represent the attenuation of WiFi signals.
	\item When the positions of transmitter (camera) and receiver (\name) are fixed, and the obstacle (user) walks in a straight line past the receiver and through the FFZ, the length of the path traversing the FFZ is related to the angle between the walking path and LOS (azimuth).
\end{itemize}
Based on the above key points, it is evident that if the user's walking speed and the distance between the transmitter and receiver are known, the azimuth angle of the wireless camera can be calculated using the time of significant CSI attenuation. 
Furthermore, an important corollary is derived:
\begin{tcolorbox}[colback=gray!25!white, size=title, breakable, boxsep=1mm, colframe=white, before={\vskip1mm}, after={\vskip0mm}]
\textbf{Corollary:} In an indoor environment, for a camera to effectively monitor an area of interest, its LOS must remain unobstructed. Therefore, if the azimuth angle of the wireless camera is known, the camera is likely located at the first obstacle encountered along that angle. 
\end{tcolorbox}
%
%
From the corollary, we know that in an indoor environment, effective localization of a wireless camera can be achieved by knowing the azimuth angle information, even without distance information. 
However, some challenges arise in practice:
\begin{itemize}
    \item Users' walking speeds are difficult to obtain.
    \item Some users may be unaware of their own sizes. 
    \item The distance between the \name device and the wireless camera is unknown.
\end{itemize}
%

\begin{figure}[t]
  \centering
  \includegraphics[width=0.78\linewidth]{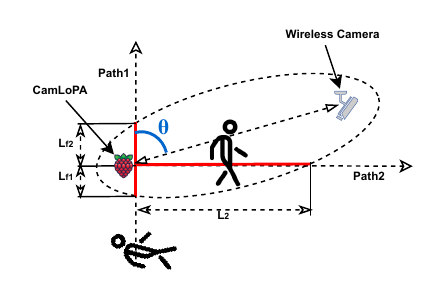}
  \vspace{-0.1in}
  \caption{The illustration of azimuth localization.}
  \label{fig:amizuth}
  \vspace{-0.22in}
\end{figure}

\name introduces the \textit{\textbf{orthogonal ratio}} to address the challenge of obtaining crucial parameters (e.g., speed and distance). As shown in Figure~\ref{fig:amizuth}, \name prompts the user to walk along two orthogonal paths, both of which pass by the \name device. In real-world environments, finding such paths is usually feasible. \name then calculates the time it takes to traverse the FFZ along each path (represented by the red lines) based on the periods of significant CSI attenuation and computes their ratio. The azimuth angle $\theta$ (the angle of the Path 1 relative to the LOS path) is estimated using a model that relates this ratio to the azimuth angle. The orthogonal ratio-based method eliminates the impact of walking speed and reduces errors due to unknown distances between devices and the user's size. 

Next, we provide a detailed explanation of the azimuth localization model based on the orthogonal ratio. As explained in Section~\ref{subsec:diffatt}, the duration of significant CSI attenuation corresponds to the time it takes for the user to traverse from entering to exiting the FFZ. Therefore, for Path 1, the walking distance that causes significant attenuation can be calculated as follows:
\begin{equation}
	L_1 = B_s + L_f,
\end{equation}
where $B_s$ and $L_f$ represent the user's body size and the length of Path 1 within the FFZ (red line in Figure~\ref{fig:amizuth}).
$L_f$ can be further divided into  $L_{f1}$, the distance from the FFZ boundary to \name, and $L_{f2}$, the distance from \name to the FFZ boundary. Combined with Equation~\ref{equ:fz}, we have the following equations:
\begin{equation}
	L_{f1}+\sqrt{d^2+L_{f1}^2-2dL_{f1}\cos\theta}-d=\frac{\lambda}{2},
\end{equation}
\begin{equation}
	L_{f2}+\sqrt{d^2+L_{f2}^2-2dL_{f1}\cos(\pi-\theta)}-d=\frac{\lambda}{2},
\end{equation}
where $d$ is the distance between $T_x$ and $R_x$. Treating $L_{f1}$ and $L_{f2}$ as unknown, they can be solved as follows:
\begin{equation}
	L_{f1}=\frac{\lambda^2+4d\lambda}{4(2d+\lambda-2d\cos\theta)},
\end{equation}
\begin{equation}
	L_{f2}=\frac{\lambda^2+4d\lambda}{4(2d+\lambda+2d\cos\theta)}.
\end{equation}
Path 2 does not cross the entire FFZ, and thus the length of its path that perturbs the CSI is only the distance from \name to the FFZ boundary:
\begin{equation}
	L_2+\sqrt{d^2+L_2^2-2dL_2\cos(\frac{\pi}{2}-\theta)}=\frac{\lambda}{2}.
\end{equation}
Treating $L_2$ as unknown, it can be solved as follows:
\begin{equation}
	L_2=\frac{\lambda^2+4d\lambda}{4(2d+\lambda-2d\sin\theta)}.
\end{equation}
The orthogonal ratio is calculated as:
\begin{equation}
\label{equ:ort}
\small
\begin{aligned}
	R_o&=\frac{T_1}{T_2}=\frac{T_1v_s}{T_2v_s}=\frac{L_1}{L_2}=\frac{4B_s(2d+\lambda-2d\sin\theta)}{\lambda^2+4d\lambda}\\&+\frac{4(2d+\lambda-2d\sin\theta)}{4(2d+\lambda-2d\cos\theta)}+\frac{4(2d+\lambda-2d\sin\theta)}{4(2d+\lambda-2d\cos\theta)}\\&=\frac{4B_s(2d+\lambda-2d\sin\theta)}{\lambda^2+4d\lambda}+\frac{8(2d+\lambda)(2d+\lambda-2d\sin\theta)}{(2d+\lambda)^2-(2d\cos\theta)^2},
\end{aligned}
\end{equation}
where $T_1$ and $T_2$ are the periods during which the user's movement along Paths 1 and 2 causes significant CSI attenuation, and $v_s$ is the user's walking speed. By taking the ratio, the influence of the speed can be eliminated. After obtaining $R_o$, the Newton-Raphson method can be used to solve for $\theta$. 

\begin{figure*}[t]
\hspace{4.8em}
\subfigure[The variations of $L_1$.]{
    \centering
    \includegraphics[width=0.4\textwidth]{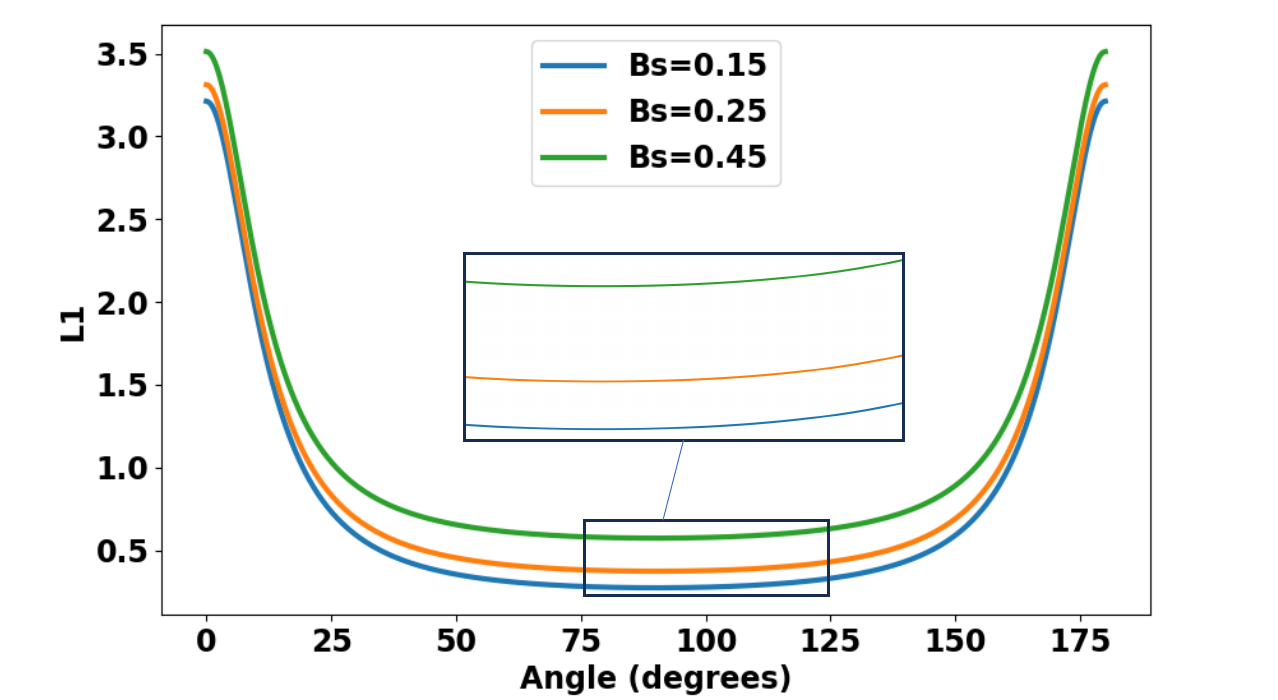}
    \vspace{-0.15in}
    \label{fig:bsl11}
    \vspace{-0.15in}}
\subfigure[The variations of $R_o$.]{
    \centering
    \includegraphics[width=0.4\textwidth]{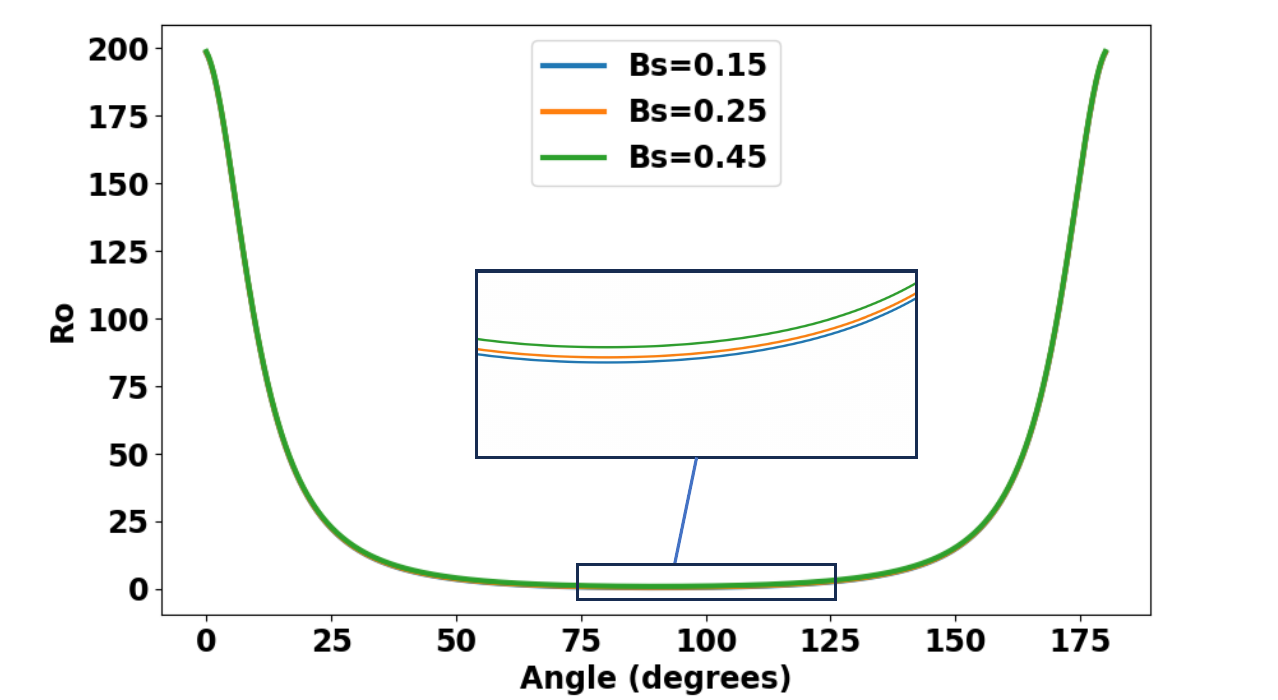}
    \vspace{-0.15in}
   \label{fig:bsro1}
   \vspace{-0.15in}}
	\caption{The variations of $L_1$ and $R_o$ relative to $\theta$ with $B_s$ changes.}
	\label{fig:bs}
	\vspace{-0.12in}
\end{figure*}

\begin{figure*}[t]
	\centering
\subfigure[The variations of $L_1$.]{
    \centering
    \includegraphics[width=0.4\textwidth]{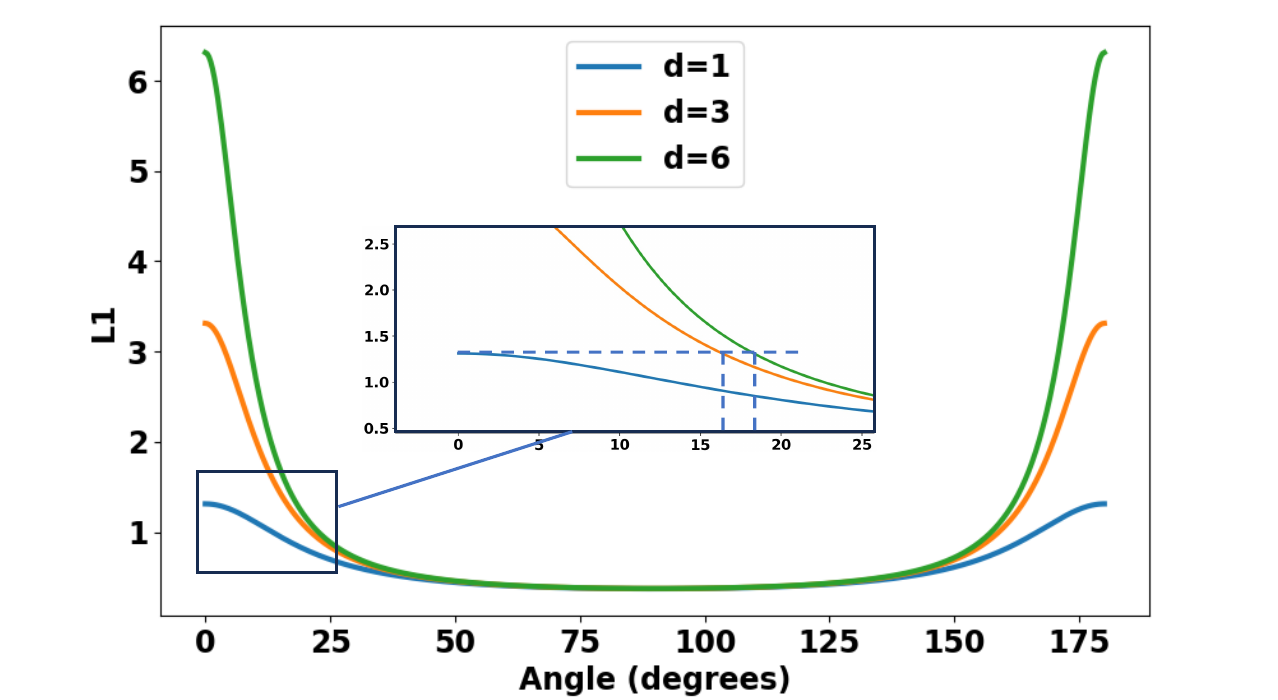}
    \vspace{-0.15in}
    \label{fig:dl11}
    \vspace{-0.15in}}
\subfigure[The variations of $R_o$.]{
    \centering
    \includegraphics[width=0.4\textwidth]{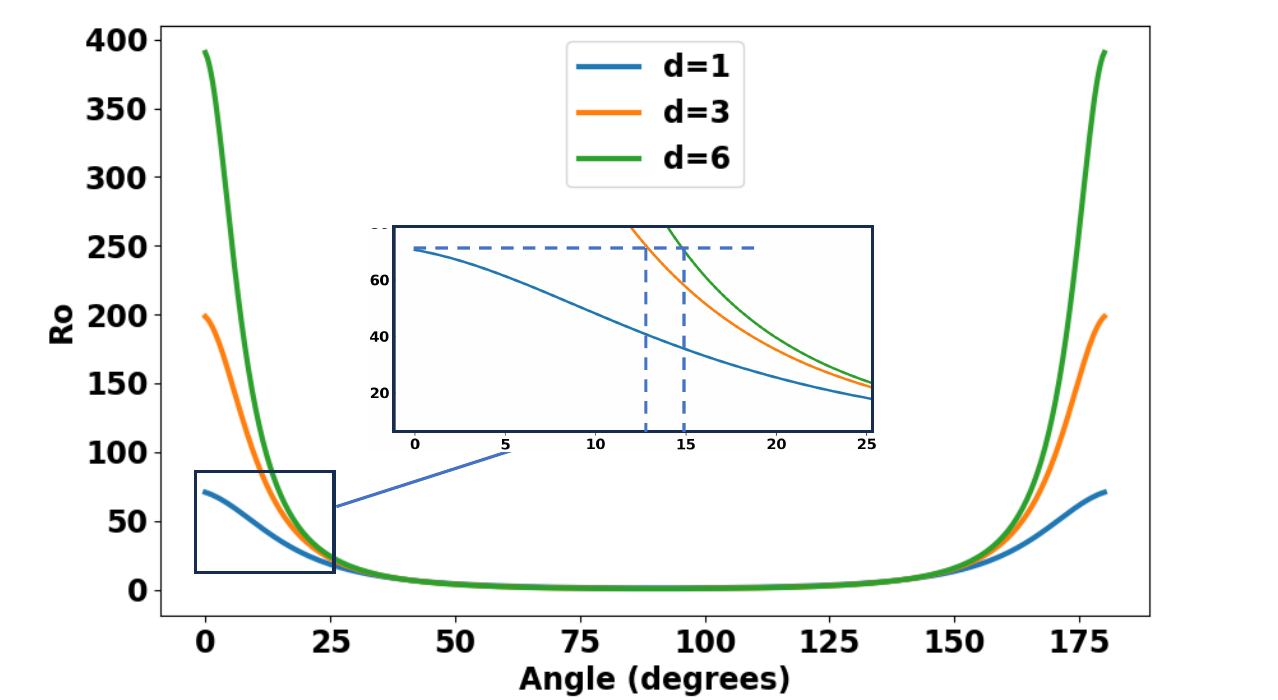}
    \vspace{-0.15in}
   \label{fig:dro1}
   \vspace{-0.15in}}
	\caption{The variations of $L_1$ and $R_o$ relative to $\theta$ with $d$ changes.}
	\label{fig:d}
	\vspace{-0.15in}
\end{figure*}

Next, we analyze the errors introduced by setting fixed values of $B_s$ and $d$. We conducted an analysis of the $L_1$-$\theta$ and $R_o$-$\theta$ relationship models separately. Figure~\ref{fig:bs} shows the variations of $L_1$ and $R_o$ relative to the azimuth angle $\theta$ for $B_s =$ 0.15, 0.25, and 0.45, which are reasonable based on common sense. It can be observed that the error caused by $B_s$ is more pronounced near $\theta = 90^o$. The error in the $L_1$-based method due to changes in $B_s$ is significant, while the $R_o$-based method effectively mitigates the error caused by the variations of $B_s$. Figure~\ref{fig:d} illustrates the variations of $L_1$ and $R_o$ relative to the azimuth angle $\theta$ for $d =$ 1, 3, and 6, which are plausible ranges for indoor wireless camera deployment. It can be observed that the error caused by $d$ is more significant around 0/180$^o$. Compared to the $L_1$-based approach (with an theoretical maximum error approaching 20$^o$), the theoretical maximum error of $R_o$ (15$^o$) is more advantageous. Furthermore, the variations in the walking speed due to different users' habits can introduce greater errors in the $L_1$-based scheme. It is clear that the orthogonal ratio-based scheme employed by \name nearly eliminates the bias caused by unknown speeds and user body sizes while minimizing the errors due to the unknown distance between the transmitter and receiver. Even under the condition of maximum theoretical error, the localization results remain highly practical in real indoor environments due to the limited number of potential hiding spots for wireless cameras. Due to the superiority of the orthogonal ratio strategy, in this paper, \name sets $d=3$ and $B_s=0.25$ as fixed values according to realistic scenarios, and users walk for 10 seconds along each path. 

\subsection{Quadrant Determination}

From Figures~\ref{fig:bs} and~\ref{fig:d} (i.e., $R_o$ leading to two possible values of $\theta$), we can also observe that the predicted $\theta$ using $R_o$ has two possible values, making it impossible to determine whether the camera is in the first or second quadrant. Therefore, further quadrant determination is necessary.

To achieve quadrant determination, \name prompts the user to walk again in the same direction as Path 1 for 10 seconds, but starting from a position in front of the \name device. The quadrant can then be determined based on changes in the CSI. The rationale is that if the wireless camera is located in the first quadrant, the user standing at the starting position will block the LOS signal between the two devices, causing significant signal variations due to the diffraction effect when the user moves. Conversely, if the wireless camera is behind the user, the user's movement will only cause signal fluctuations due to reflection. Specifically, \name determines the quadrant as follows:
\begin{equation}
\label{equ:qufind}
	\mathbf{Q_{mac}} =\begin{cases}
	\mathbf{2}  & \text{ if } \frac{\max(CSI_3)}{\min(CSI_3)}<T_q*\frac{\max(CSI_1)}{\min(CSI_1)},  \\
	\mathbf{1}  & \text{ else }.
	\end{cases}
\end{equation}
Equation~\ref{equ:qufind} means that if the extent of the CSI fluctuation caused by Path 3 is less than $T_q$ times the extent of the CSI fluctuation caused by Path 1, the camera is determined to be in the second quadrant; otherwise, it is in the first quadrant.

Since movement within the range of 180-360$^o$ does not cross the LOS, \name can only locate devices within the range of 0-180$^o$. However, in real-world environments, the user's available space is usually near walls, thus a single measurement by \name remains highly useful. If the condition of moving near walls is not met, \name requires two measurements.

\section{Implementation and Evaluation}
We implemented \name in multiple rooms and diverse hidden wireless cameras, and this section presents the implementation and evaluation details of \name.

\begin{figure}[t]
  \centering
  \includegraphics[width=0.75\linewidth]{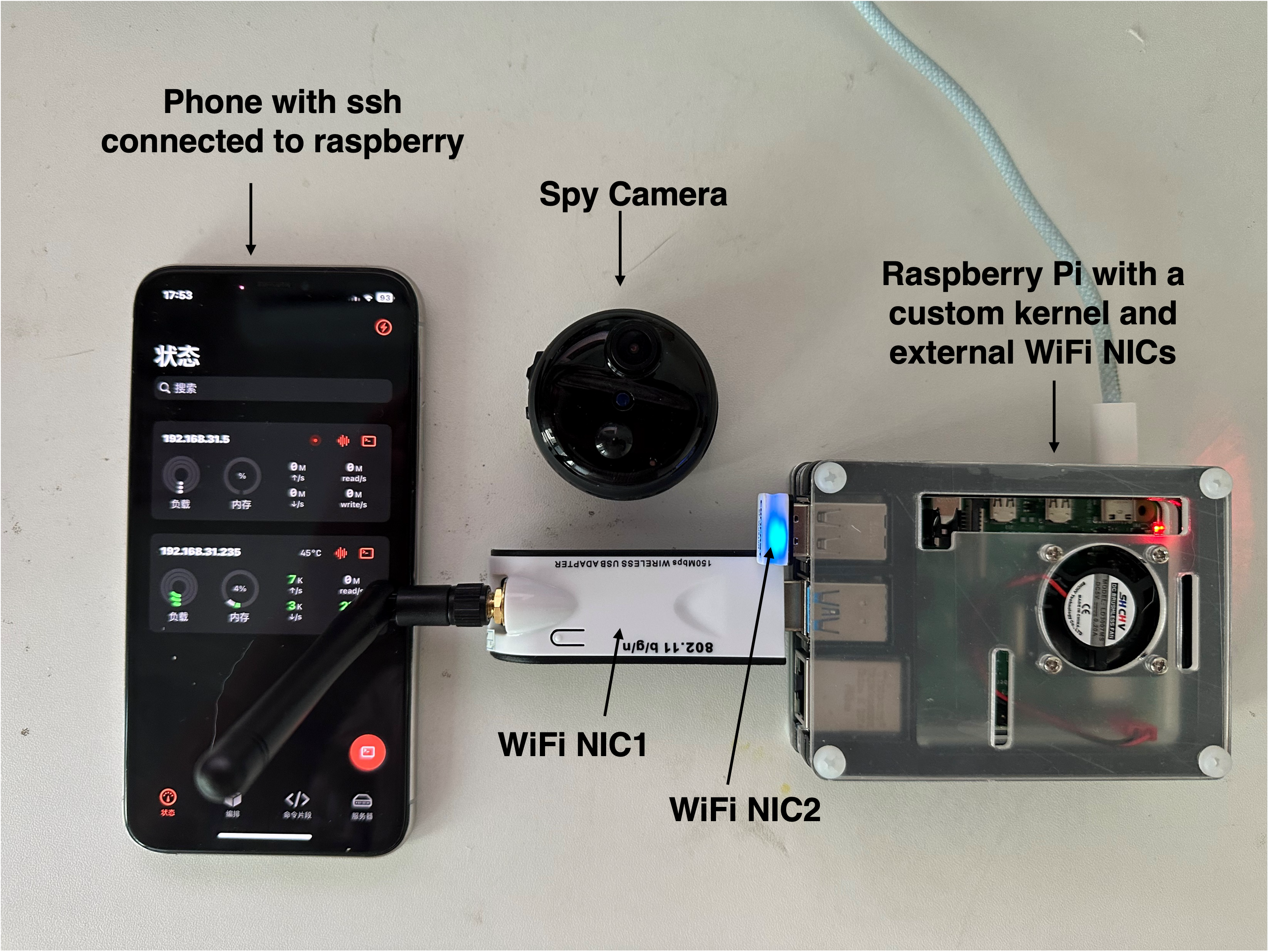}
  \caption{The prototype of \name.}
  \label{fig:prototype}
  \vspace{-0.15in}
\end{figure}

\begin{figure*}[t]
\centering
\subfigure[Room1]{
    \centering
    \includegraphics[width=0.3\textwidth]{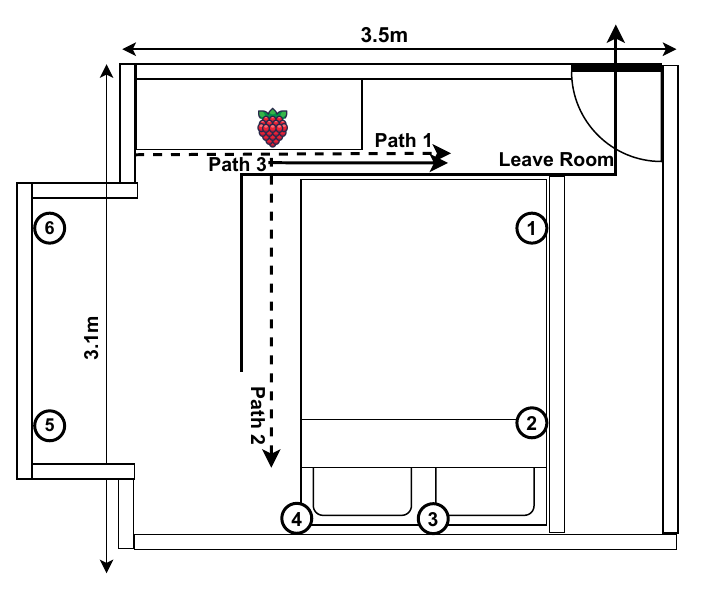}
    \vspace{-0.05in}
    \label{fig:room1}
}
\vspace{-0.03in}
\subfigure[Room2]{
    \centering
    \includegraphics[width=0.22\textwidth]{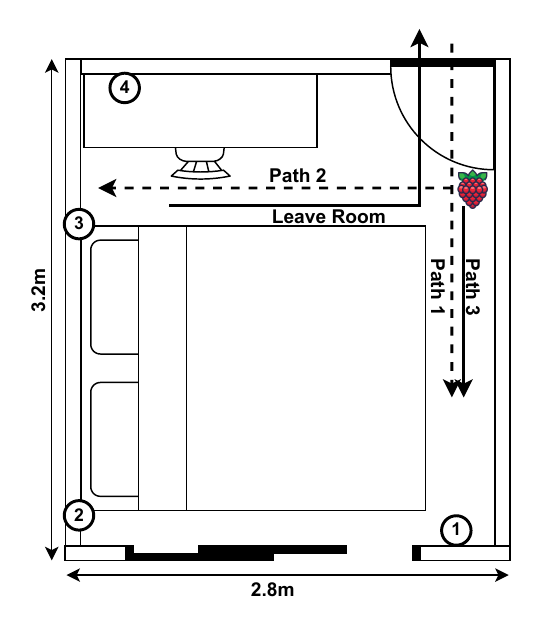}
    \vspace{-0.05in}
   \label{fig:room2}
}
\vspace{-0.03in}
\subfigure[Room3]{
    \centering
    \includegraphics[width=0.36\textwidth]{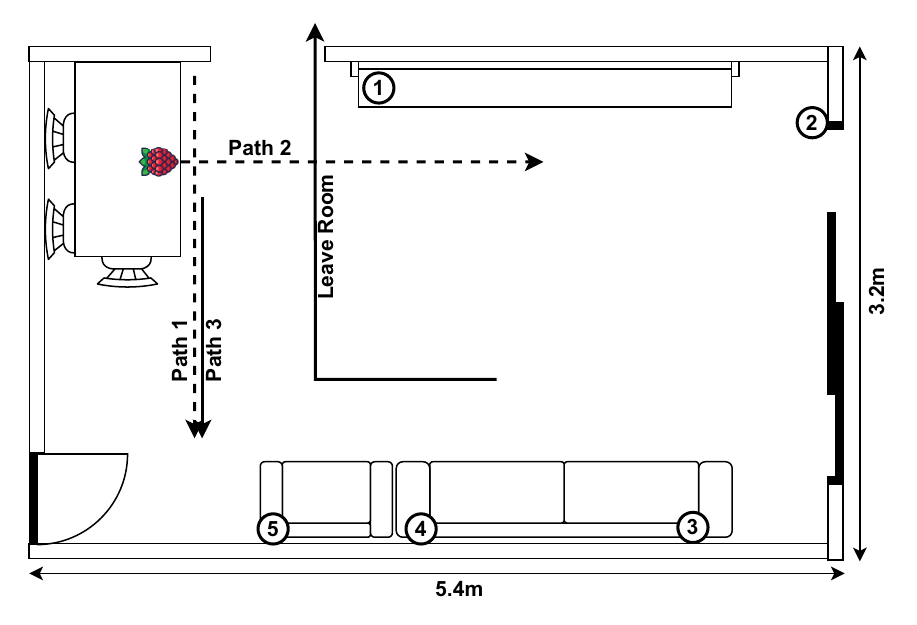}
    \vspace{-0.05in}
   \label{fig:room3}
}
\vspace{-0.03in}
	\caption{The layout of three rooms.}
	\label{fig:room}
	\vspace{-0.15in}
\end{figure*}

\subsection{Prototype}

\name needs to sniff 802.11 packets in the air and obtain CSI. Currently, most mobile devices require special permissions to perform sniffing. Otherwise, due to the closed-source nature of wireless network card manufacturers, extracting CSI is only possible with certain network cards. However, acquiring this data poses no technical challenge but only involves permission issues. To ensure system applicability, we did not implement \name on specific phone or computer platforms which can extract CSI. Instead, we chose the open-source and low-cost COTS device, Raspberry Pi, as the platform for \name.

The prototype of \name is shown in Figure~\ref{fig:prototype}. The Raspberry Pi uses its built-in wireless NIC with the nexmon tool~\cite{gringoli2019free} to modify the kernel for CSI extraction. However, the modified driver cannot sniff normal 802.11 packets, therefore we set up an external network card (NIC1) with monitoring capabilities to sniff 802.11 packets. NIC2 is a standard wireless network card used for communication between the \name device and the user's smartphone. The user's smartphone can receive prompts and localization results from \name via SSH tools.

\subsection{Experimental Setup}

\begin{table}[t]
\centering
\caption{Cameras used in experiments.}
\begin{tabular}{c|c|c} \hline
 Camera & Abbreviation & Cost(\$) \\ \cline{1-3}
 XiaoMi Cloud Camera2 & Mi & 24.5\\ \cline{1-3}
 XiaoYi Smart Camera Y4 & Yi & 20.4\\ \cline{1-3}
 EZVIZ C2C & C2C & 24.5\\ \cline{1-3}
 360 Cloud Camera 8Pro & 360 & 24.5\\ \cline{1-3}
 V380 Camera & V380 & 13.6\\ \cline{1-3}
 Guangchun Mini Camera & Gc & 31.4\\ \cline{1-3}
 HiLEME Mini Camera & Hi & 18.4\\ \cline{1-3}
\end{tabular}
\label{tab:camdev}
\vspace{-0.2in}
\end{table}

We evaluated the performance of \name on seven different wireless cameras, as listed in Table~\ref{tab:camdev}. All devices were purchased from shopping platforms, and the cameras were connected to a 2.4GHz WiFi network for data transmission. The experiments were conducted in three real-life rooms. Since the experiments were carried out in actual home environments and lasted for an extended period, only the residents participated to protect privacy. The validation experiments lasted for two months in total.

The layout of three rooms are shown in Figure~\ref{fig:room}. Rooms 1 and 2 (Figures~\ref{fig:room1} and~\ref{fig:room2}) are bedrooms, while room 3 is a living room (Figure~\ref{fig:room3}). In real environments, private spaces like bedrooms and hotel rooms have limited activity space, restricting the feasibility of previous methods that rely on extensive indoor scanning. In each room, we select several potential locations suitable for monitoring the entire room to place the cameras for the experiments. The azimuths (path 1 as x-axis) of each point in room 1 are 28.61$^o$, 42.27$^o$, 60.28$^o$, 88.54$^o$, 130.1$^o$, and 157.73 $^o$, in room 2 are 4.86$^o$, 51.34$^o$, 69.44$^o$, and 103.52 $^o$, and in room 3 are 110.94$^o$, 92.37$^o$, 61.34$^o$, 47.13$^o$, and 30.69 $^o$.

\begin{figure}[t]
  \centering
  \includegraphics[width=0.9\linewidth]{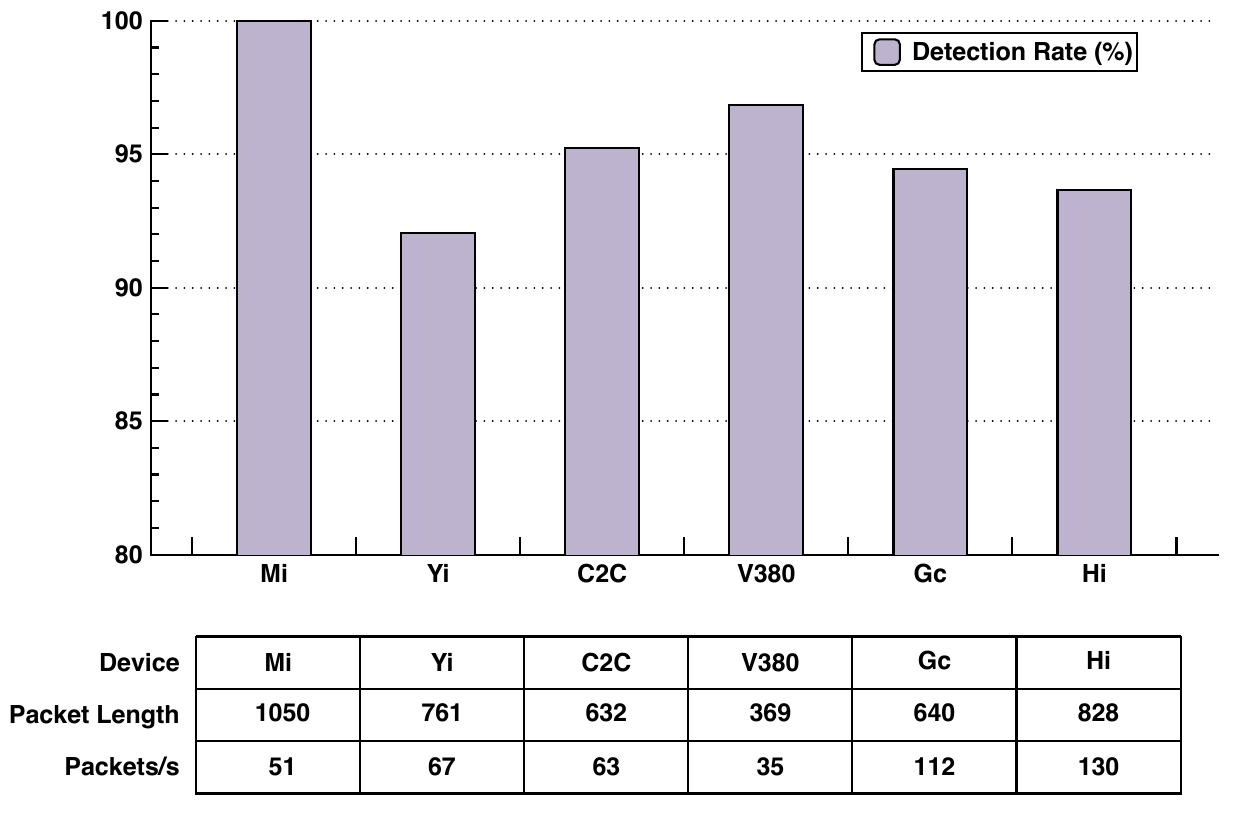}
  \vspace{-0.15in}
  \caption{Snooping camera detection performance.}
  \label{fig:snoopde}
  \vspace{-0.23in}
\end{figure}

As shown in Figure~\ref{fig:snoopde}, the cameras we used have an average QoS data packet length ranging from 369 to 1050 bytes during video stream uploads, with upload speeds ranging from 35 to 130 packets per second. Therefore, in our experiments, $T_s$ and $T_l$ are set to 300 bytes and 150 packets (30 packets * 5 seconds), respectively. The $T_q$ for quadrant localization is empirically set to 0.6. For camera detection, we use the success rate of detecting snooping cameras as the evaluation metric, while for localization, we use the azimuth angle error as the evaluation metric.

\begin{figure*}[t]
	\centering
\subfigure[4.11$^o$, path1.]{
    \centering
    \includegraphics[width=0.225\textwidth]{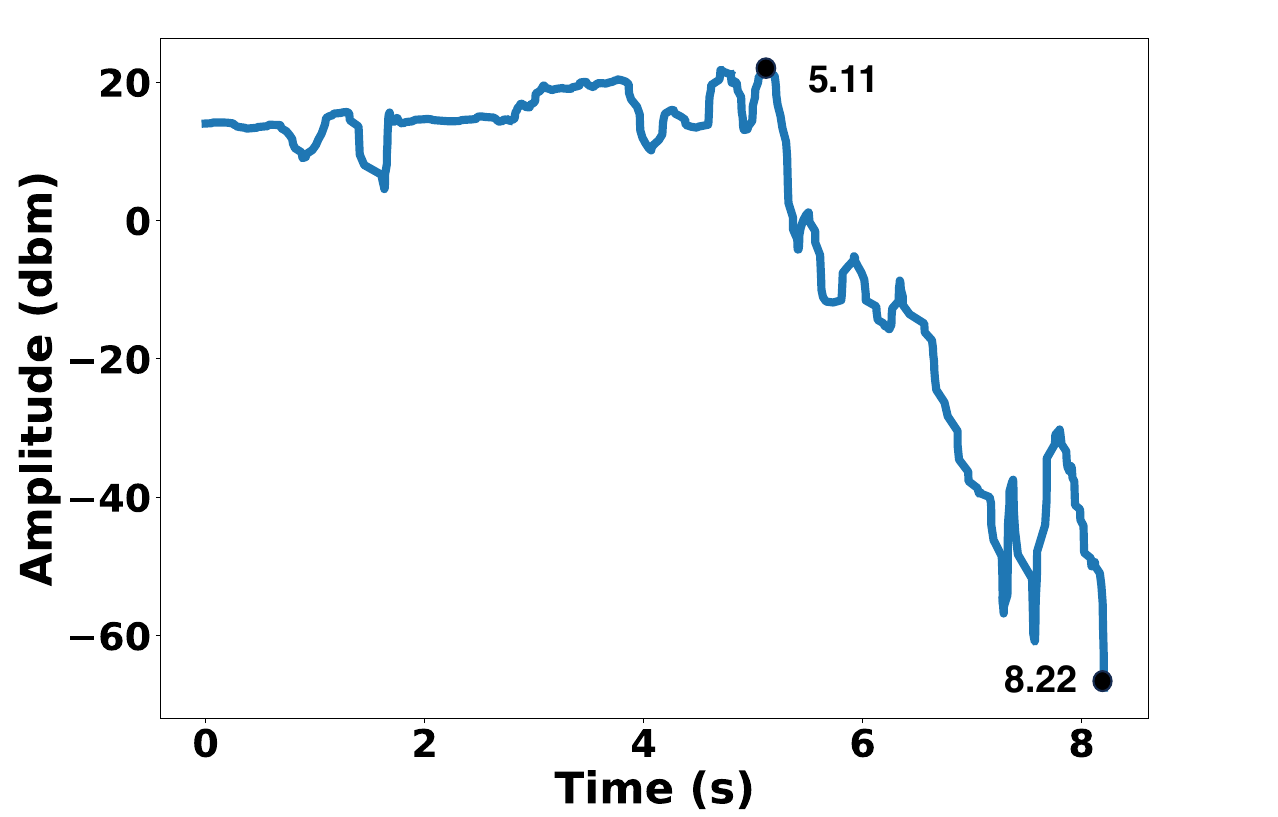}
    \label{fig:01}
    \vspace{-0.1in}}
\vspace{-0.02in}
\subfigure[28.61$^o$, path1.]{
    \centering
    \includegraphics[width=0.225\textwidth]{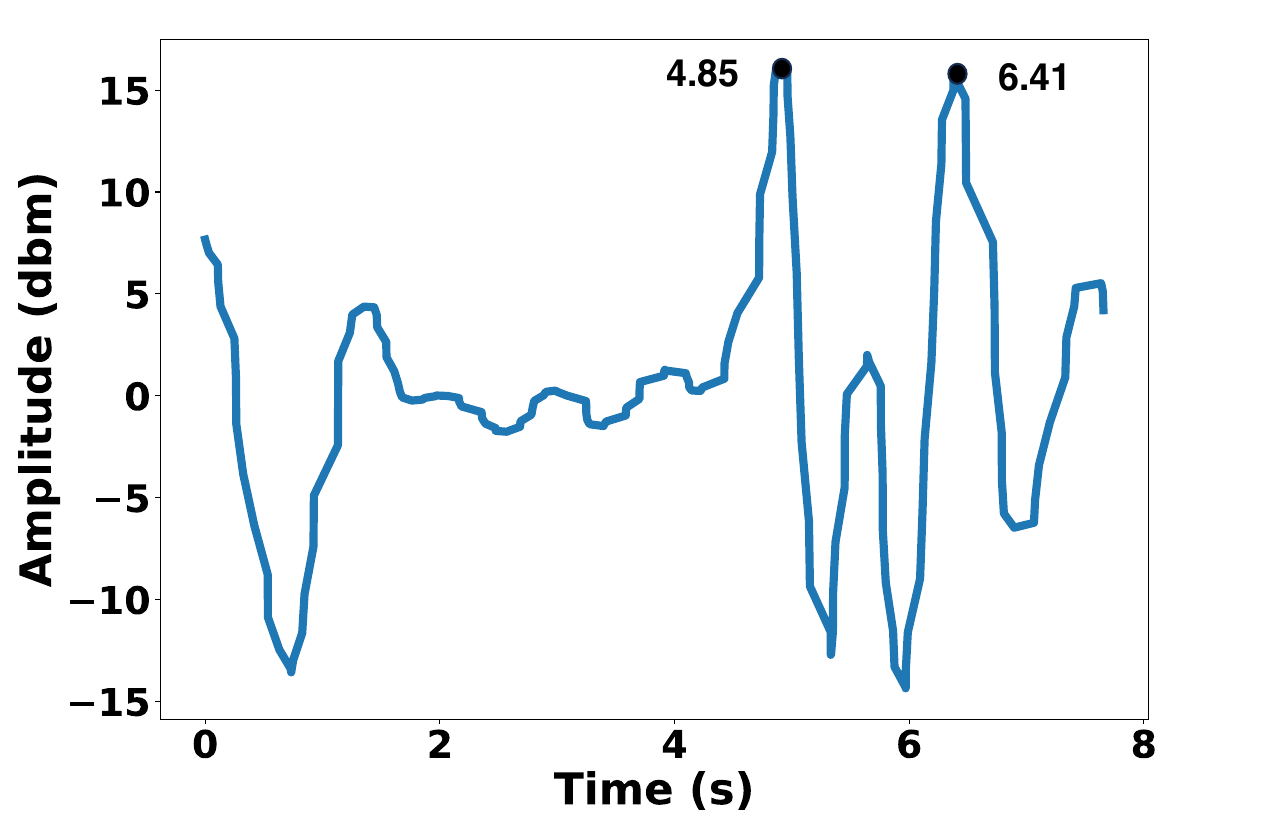}
   \label{fig:21}
   \vspace{-0.1in}
}
\vspace{-0.02in}
\subfigure[60.28$^o$, path1.]{
    \centering
    \includegraphics[width=0.225\textwidth]{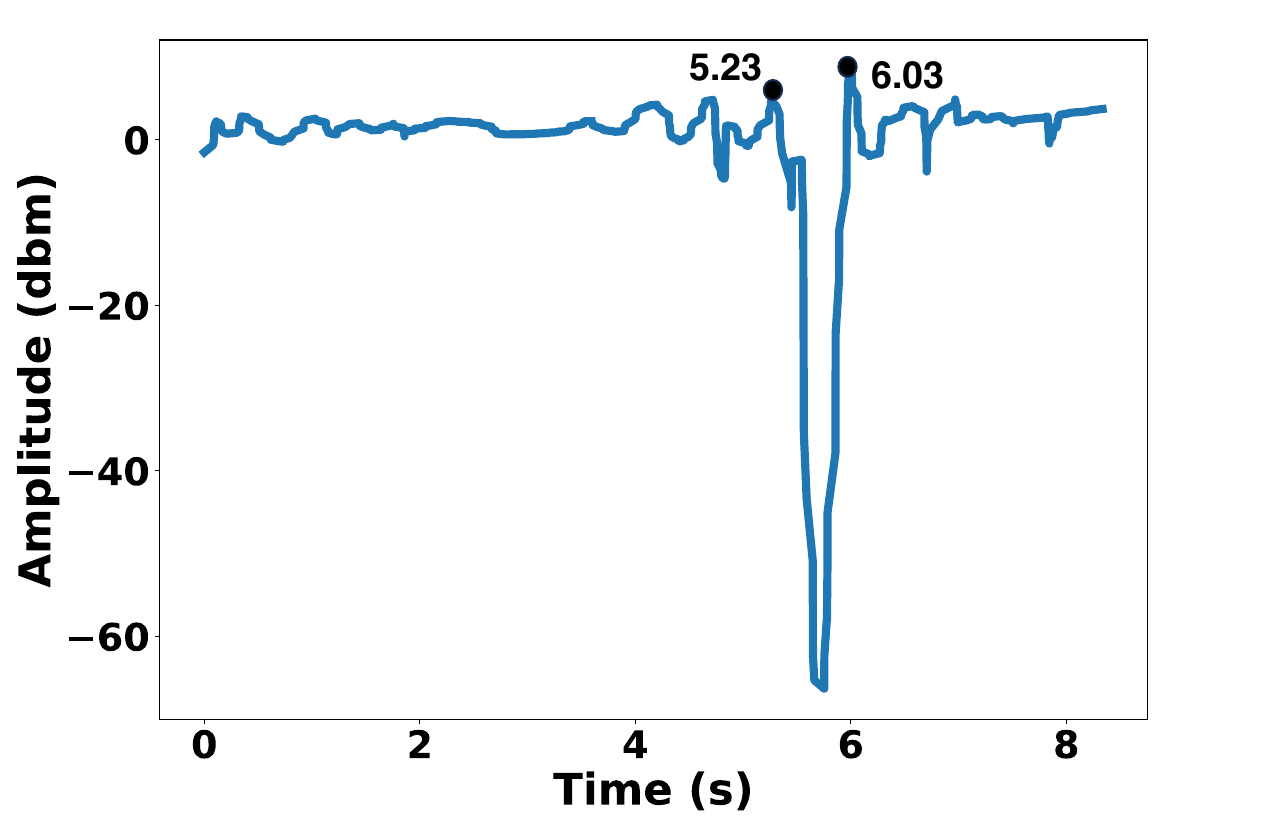}
   \label{fig:61}
   \vspace{-0.1in}
}
\vspace{-0.02in}
\subfigure[88.54$^o$, path1.]{
    \centering
    \includegraphics[width=0.225\textwidth]{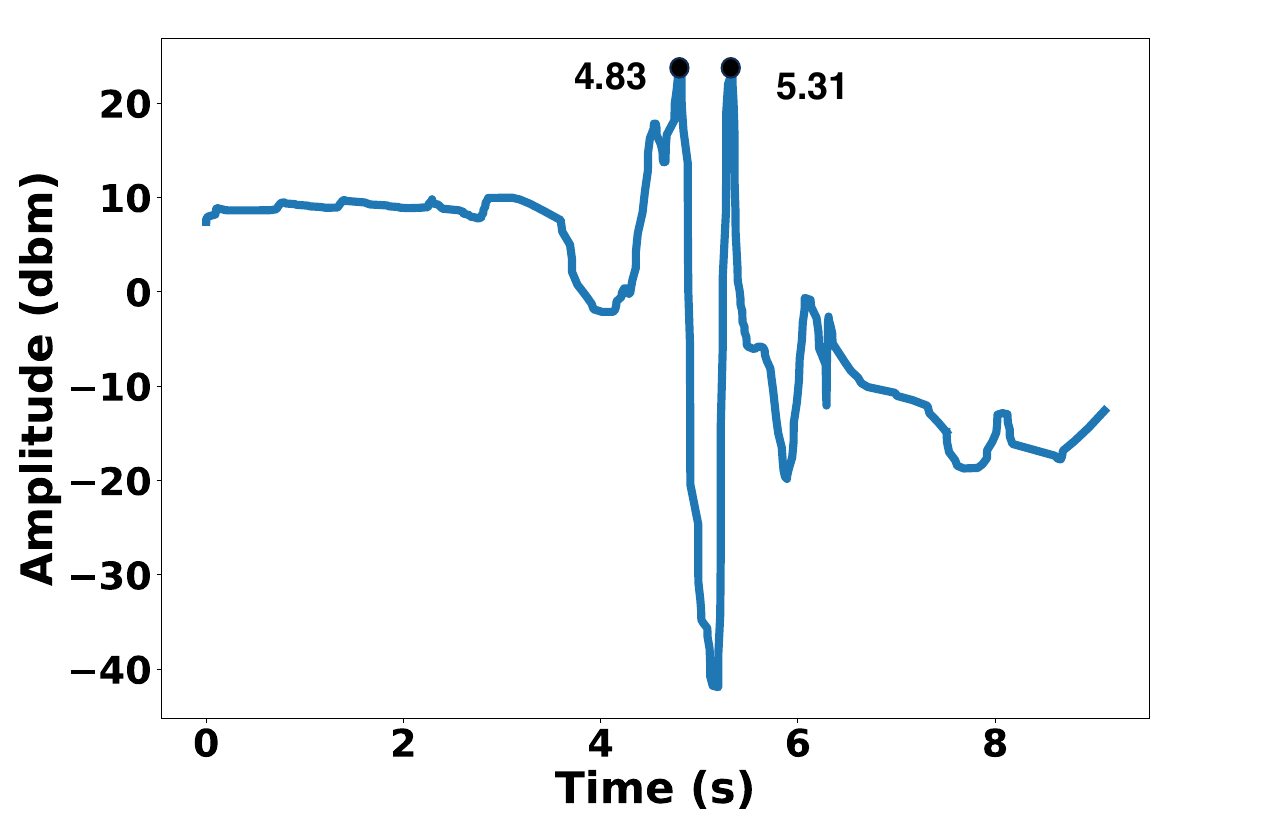}
   \label{fig:81}
   \vspace{-0.1in}
}
\vspace{-0.02in}
\subfigure[4.11$^o$, path2.]{
    \centering
    \includegraphics[width=0.225\textwidth]{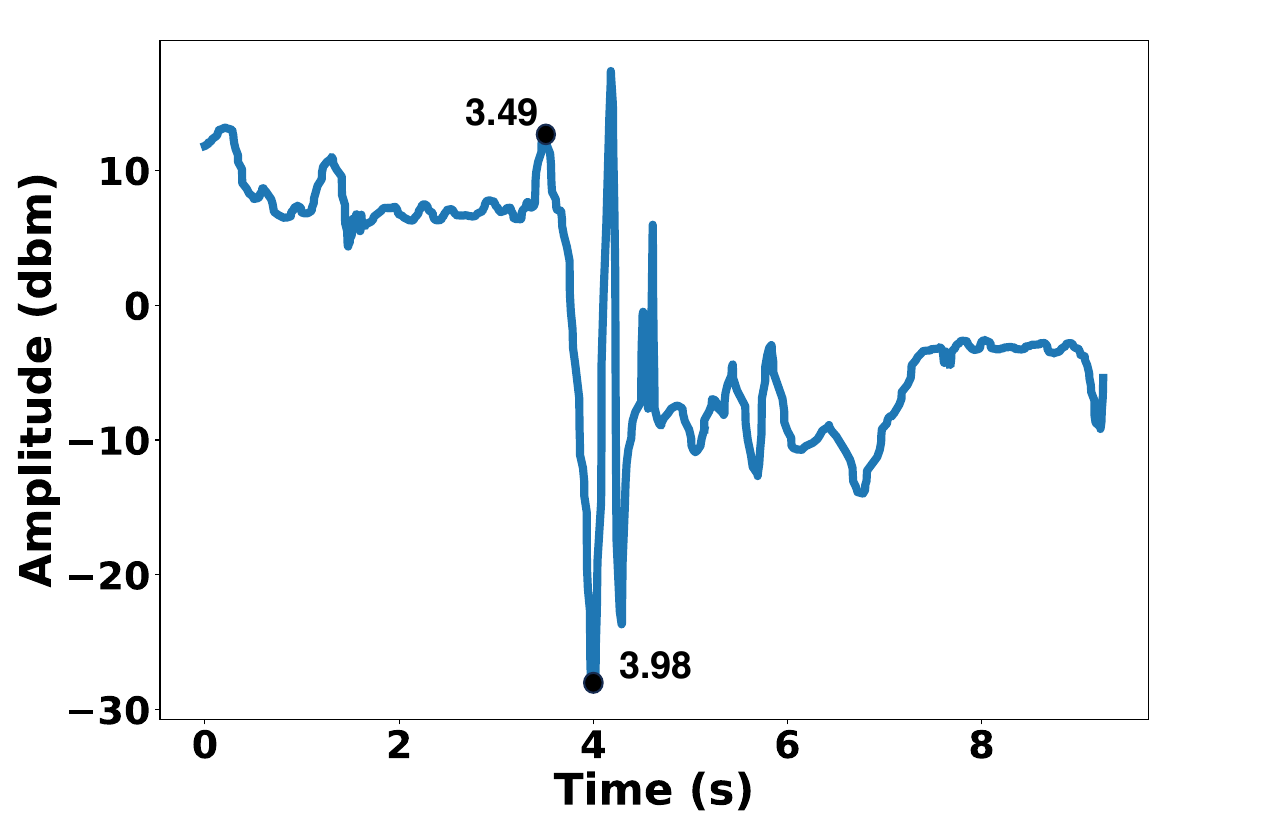}
    \label{fig:02}
    \vspace{-0.1in}
}
\vspace{-0.03in}
\subfigure[28.61$^o$, path2.]{
    \centering
    \includegraphics[width=0.225\textwidth]{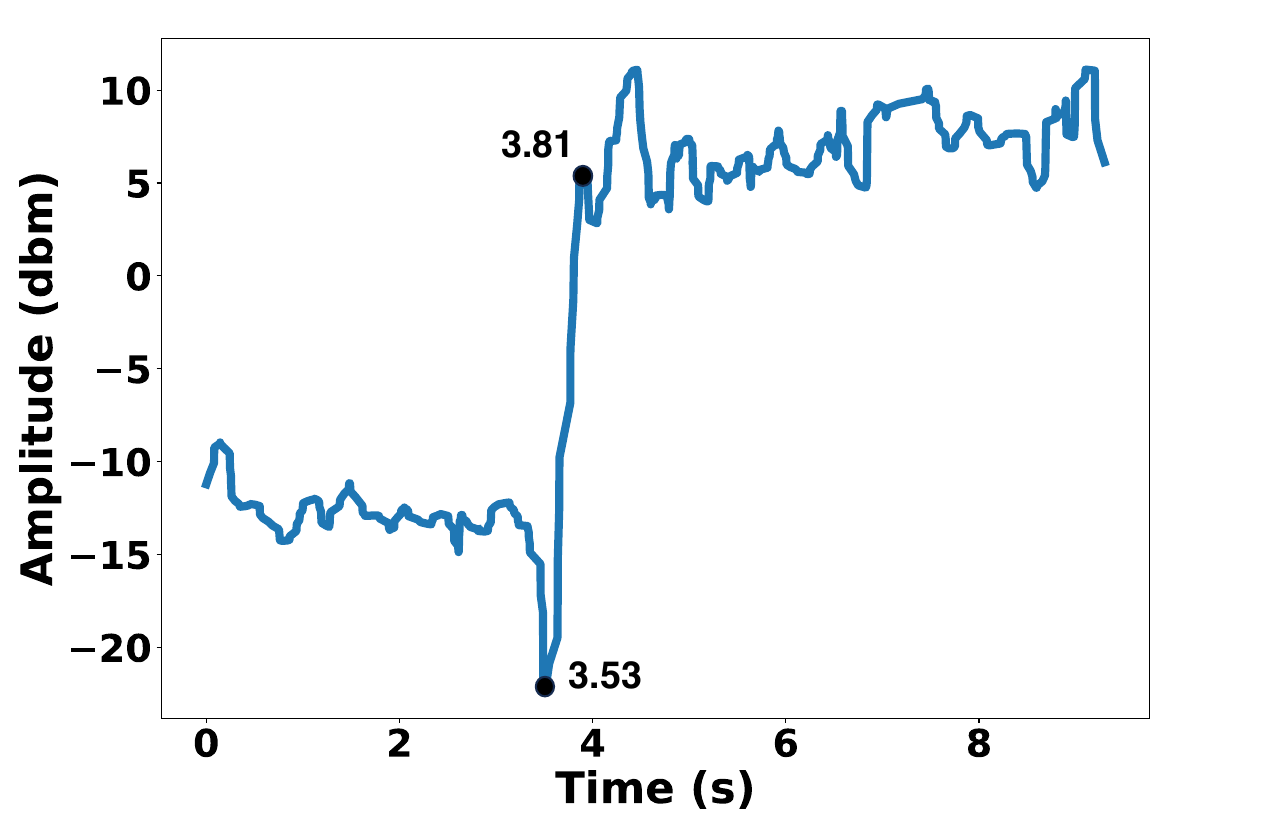}
    \label{fig:22}
    \vspace{-0.1in}
}
\vspace{-0.03in}
\subfigure[60.28$^o$, path2.]{
    \centering
    \includegraphics[width=0.225\textwidth]{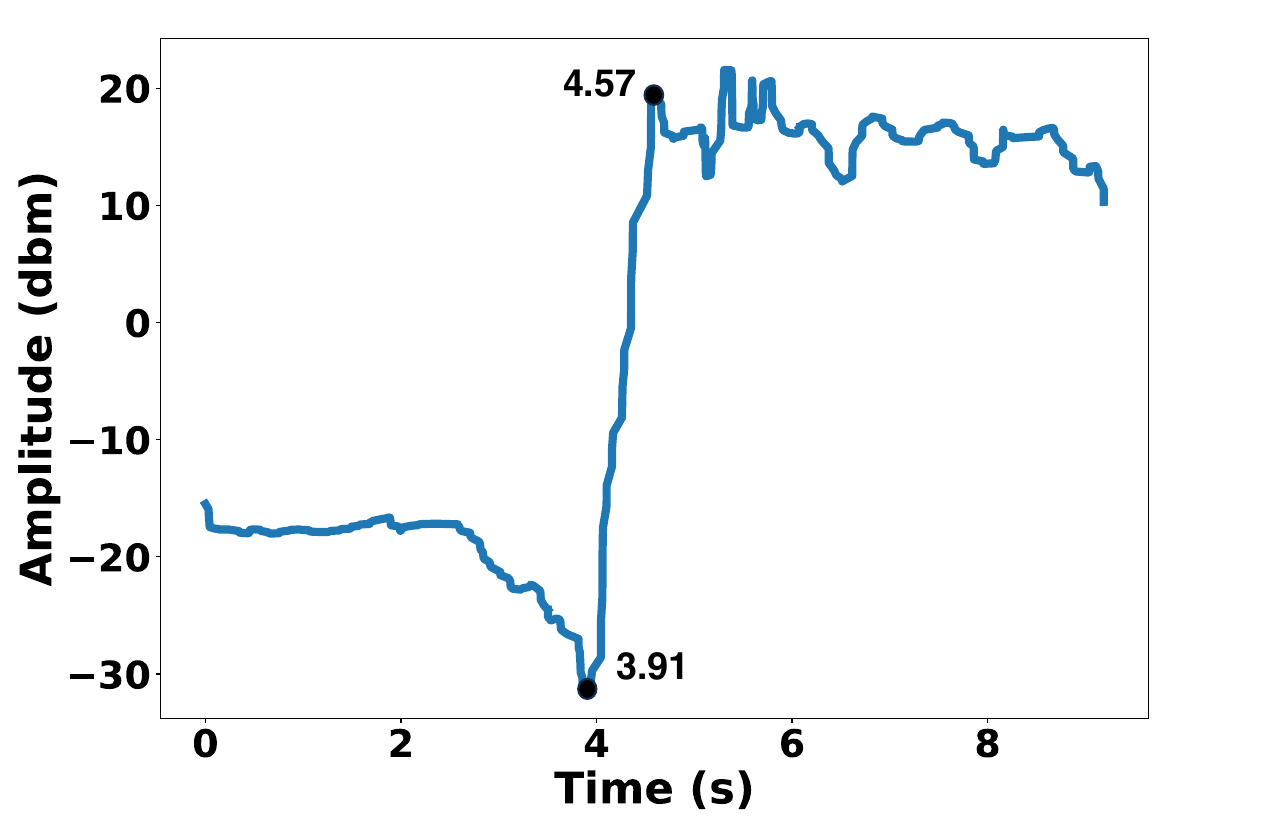}
   \label{fig:62}
   \vspace{-0.1in}
}
\vspace{-0.03in}
\subfigure[88.54$^o$, path2.]{
    \centering
    \includegraphics[width=0.225\textwidth]{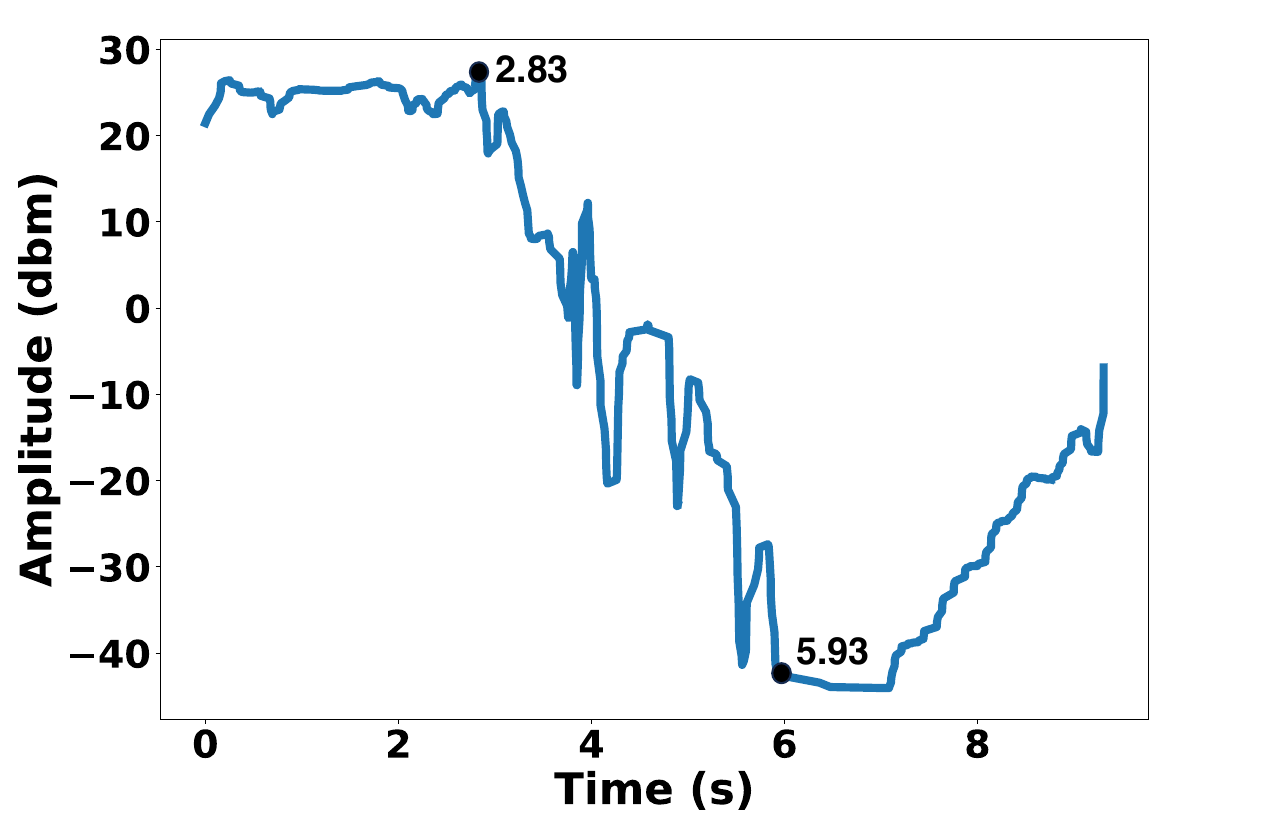}
   \label{fig:82}
   \vspace{-0.1in}
}
\vspace{-0.03in}
	\caption{The variation in CSI amplitude due to user activity when the wireless camera is positioned at different azimuth angles. The black dots represent the start and end points of significant CSI fluctuations for each path. By dividing the duration of significant attenuation of path 1 by that of path 2, we obtain $R_o$, which is then used to calculate $\theta$ according to Equation~\ref{equ:ort}. In (c) and (g), $R_o$ is calculated as $\frac{0.8}{0.66} = 1.21$, and substituting this into Equation~\ref{equ:ort} yields $\theta = 72.18^\circ$. The calculations for the others follow the same procedure.}
	\label{fig:csiana}
	\vspace{-0.19in}
\end{figure*}

\begin{figure}[t]
\centering
\subfigure[28.61$^o$, path3.]{
    \centering
    \includegraphics[width=0.36\textwidth]{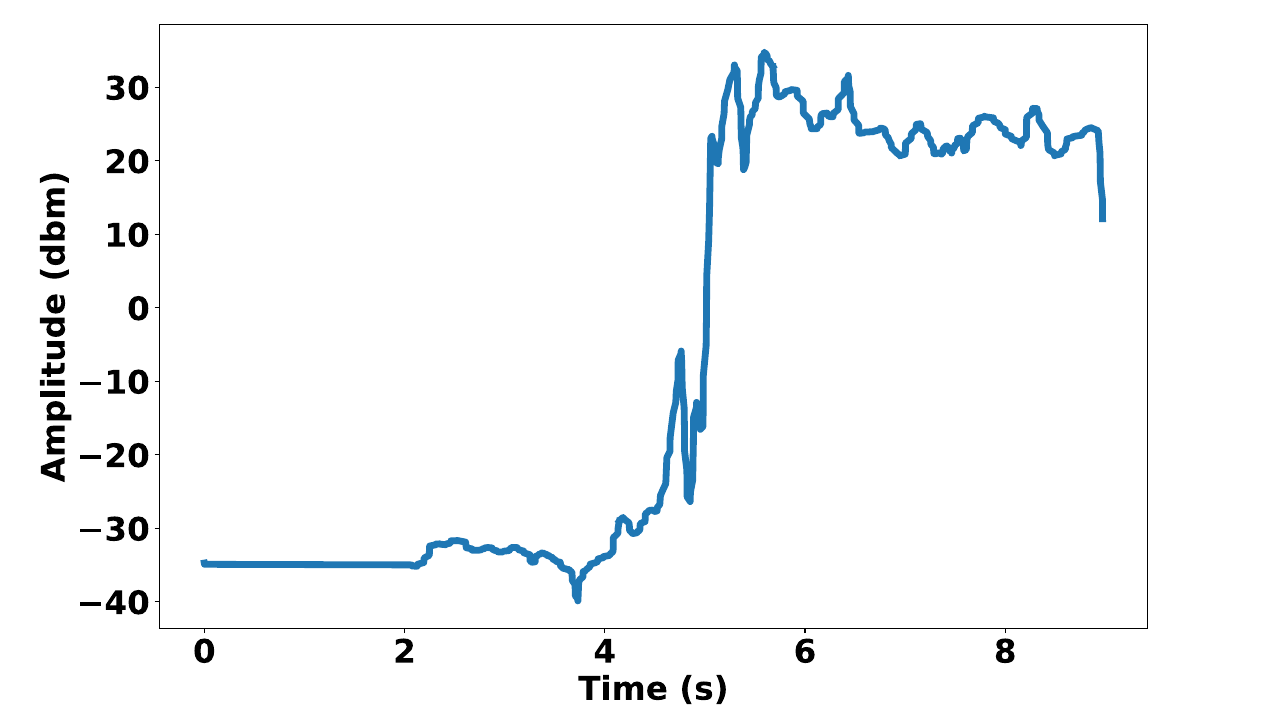}
    \label{fig:23}
}
\vspace{-0.06in}
\subfigure[130.1$^o$, path3.]{
    \centering
    \includegraphics[width=0.36\textwidth]{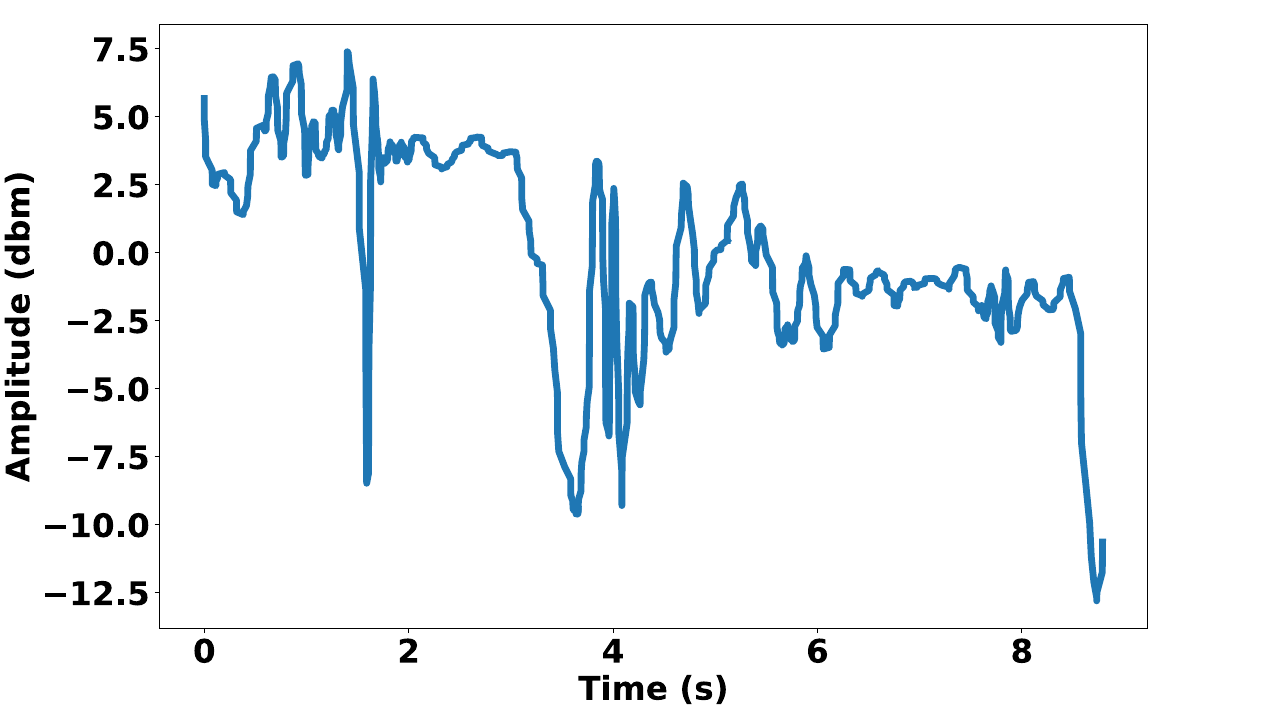}
   \label{fig:133}
}
\vspace{-0.0in}
	\caption{The variation in CSI amplitude due to user activity when the wireless camera is positioned at different quadrant. When the camera is located in the first quadrant (a), the user's starting position blocks the LOS, resulting in significant CSI fluctuations during movement. In contrast, when the camera is located in the second quadrant (b), the user does not block the LOS, leading to minor CSI fluctuations.}
	\label{fig:qualo}
	\vspace{-0.25in}
\end{figure}

\subsection{CSI Analysis and Algorithm Implementation}
\label{subsec:csi}

In this section, we analyze the mapping relationship between the CSI influenced by user activity and the angle of the wireless camera. Furthermore, we elaborate on the design of the algorithm for extracting attenuation time from the CSI.

The variation in CSI amplitude due to user activity when the wireless camera is positioned at different azimuth angles is shown in Figure~\ref{fig:csiana}. It can be observed that the CSI amplitude variation is significantly influenced by the azimuth angle of the wireless camera relative to \name. Generally, the larger the angle, the shorter the duration of significant fluctuations in CSI from Path 1 (CSI 1), while the duration of significant fluctuations in CSI from Path 2 (CSI 2) increases. These experimental results validate the feasibility of the azimuth localization scheme proposed by \name. Additionally, we made several other observations:
\begin{itemize}
	\item The fluctuation duration of CSI 2 may not accurately reflect the actual path length causing the fluctuation, as it takes time for the user to accelerate from a stationary state to walking.
	\item When the angle is too small (0 degrees) or too large (90 degrees), the calculated $R_o$ significantly deviates from the theoretical $R_o$. This is due to the limited indoor walking space usually causes the user to stop after a short distance due to obstacles, resulting in a fluctuation duration shorter than the theoretical value.\end{itemize}

To obtain the duration of significant CSI fluctuations, we use different methods for CSI 1 and CSI 2. For CSI 1, we first identify the lowest point and then use the calculated inverse to find the start and end points of the fluctuation. For CSI 2, we first calculate the mean values of the initial and later segments, then we construct a piecewise waveform where the values of the initial and later segments are equal to the calculated means. By adjusting the position of the segmentation, we find the point that best matches the waveform with CSI 2 to determine the midpoint of the fluctuation. We then calculate the inverse to identify the start and end points of the fluctuation. Additionally, based on our first observation, we scale the calculated fluctuation duration for CSI 2 to eliminate errors.  For activities that cause fluctuations exceeding a certain duration, we increase the fluctuation time to mitigate the effect noted in the second observation. As shown in Fig~\ref{fig:csiana}, CamPoLA achieves localization of cameras depolyed at different positions.

Figure~\ref{fig:qualo} shows the variations in CSI 3 (corresponding to Path 3) when the wireless camera is located in different quadrants. It is obvious that the quadrant localization scheme proposed by \name is also effective. Since CSI consists of many different subcarriers, and different subcarriers have varying sensitivities to user activity (with higher amplitudes indicating lower sensitivity), \name focuses only on the periods of significant attenuation. Therefore, we select the five subcarriers with the highest amplitudes, average them after filtering, and use this average as the final input for \name to calculate $R_o$ and the quadrant.

\begin{figure*}[t]
	\centering
\subfigure[Room1.]{
    \centering
    \includegraphics[width=0.305\textwidth]{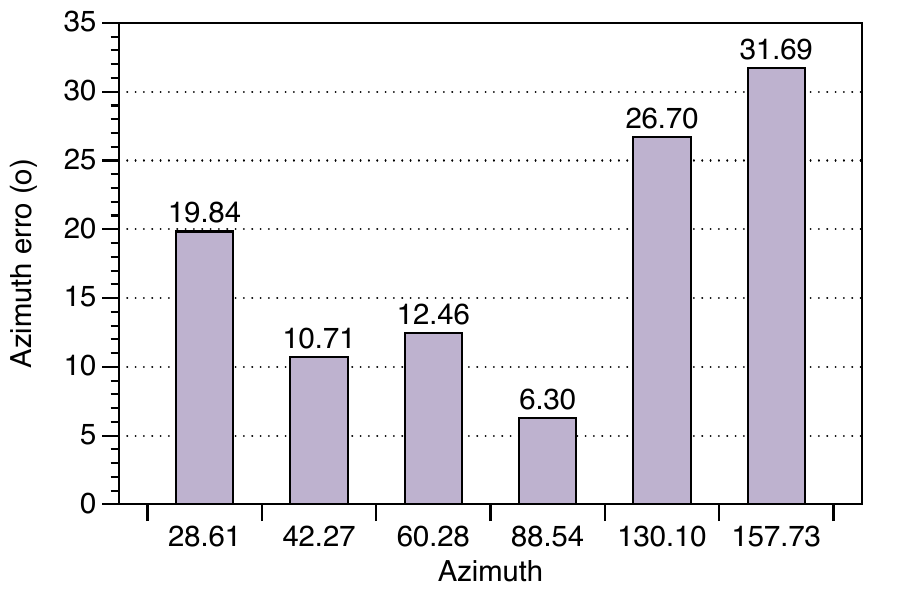}
    \vspace{-0.15in}
    \label{fig:angroom1}
}
\vspace{-0.02in}
\subfigure[Room2.]{
    \centering
    \includegraphics[width=0.305\textwidth]{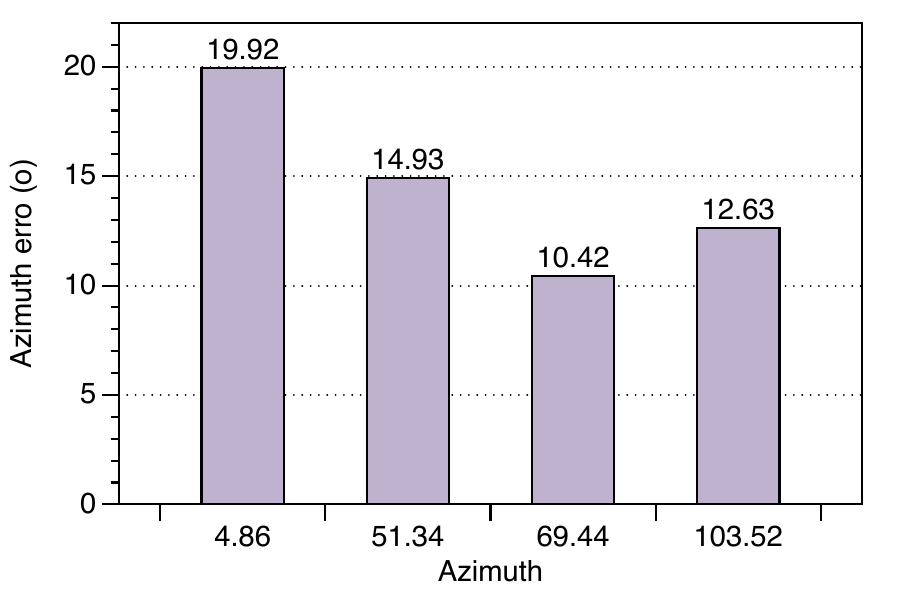}
    \vspace{-0.15in}
   \label{fig:angroom2}
}
\vspace{-0.02in}
\subfigure[Room3.]{
    \centering
    \includegraphics[width=0.305\textwidth]{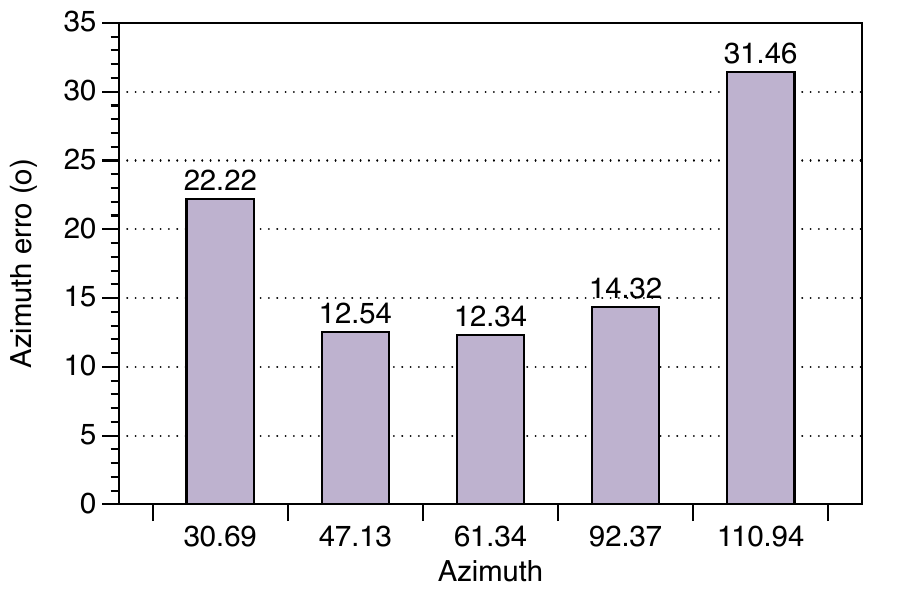}
    \vspace{-0.15in}
   \label{fig:angroom3}
}
\vspace{-0.02in}
	\caption{Localization results of hidden cameras deployed at different positions.}
	\label{fig:angroom}
	\vspace{-0.22in}
\end{figure*}

\subsection{Performance of Wireless Camera Detection}

\name detects wireless cameras monitoring the current area by first identifying suspicious devices, prompting the user to leave the room, and monitoring throughput changes to detect snooping hidden wireless cameras. \name achieves an 84.35\% success rate in identifying suspicious wireless cameras across all devices. The probability of identifying the 360 camera as a suspicious device is 0, while the accuracy of detecting other wireless cameras as suspicious devices reaches 98.41\%. This discrepancy occurs because, during traffic sniffing, the 360 wireless camera only allows the capture of ACK Block and Request-to-Send packets, but not QoS data packets. This limitation may be due to the special data transmission methods or protocols they use, which prevent its traffic from being intercepted, thus hindering detection and previous methods based on WiFi traffic all cannot work~\cite{sharma2022lumos,singh2021always,he2021motioncompass,heo2022there}. However, the nexmon tool used by \name can still capture the CSI for the 360 camera from WiFi traffic. The snooping camera detection results are shown in Figure~\ref{fig:snoopde}. \name achieves a 95.37\% success rate in detecting snooping cameras for six types of cameras across three rooms, except for the 360 wireless camera. For devices similar to the 360 camera, we believe that wireless camera detection can still be achieved by querying the OUI of the captured Request-to-Send packet's leaked MAC address. By constructing an OUI table of all available devices using device name information from shopping platforms and MAC address lookup websites, it is possible to identify the device type. However, \name cannot determine whether the camera is monitoring the current area using this method.

\subsection{Performance of Wireless Camera Localization}

The localization results for different deployment positions in three rooms are shown in Figure~\ref{fig:angroom}, and \name achieves an average azimuth localization error of 17.24 degrees for wireless hidden cameras. \name demonstrates higher localization accuracy for cameras within the 40-90$^o$ range, while the accuracy decreases for cameras in the second quadrant or close to 0$^o$. This discrepancy arises from errors introduced by the quadrant determination scheme and path length limitations. We further discuss the specific causes of these errors in Section~\ref{sec:diss}. For cameras near 90$^o$, the algorithm described in Section~\ref{subsec:csi} causes \name to tend to output predictions close to 90$^o$, resulting in lower errors. Overall, \name achieves high localization accuracy than previous similar studies in a shorter time and with lower activity space requirements, and without the need for training. Although \name's localization results are not perfectly precise in confined indoor spaces, it significantly improve the localization resolution and reduce the search area for users compare to previous studies. However, \name still has some limitations and areas for improvement, which we will discuss further in Section~\ref{sec:diss}.

\subsection{Sensitivity Analysis}

\begin{figure}[]
  \centering
  \includegraphics[width=0.65\linewidth]{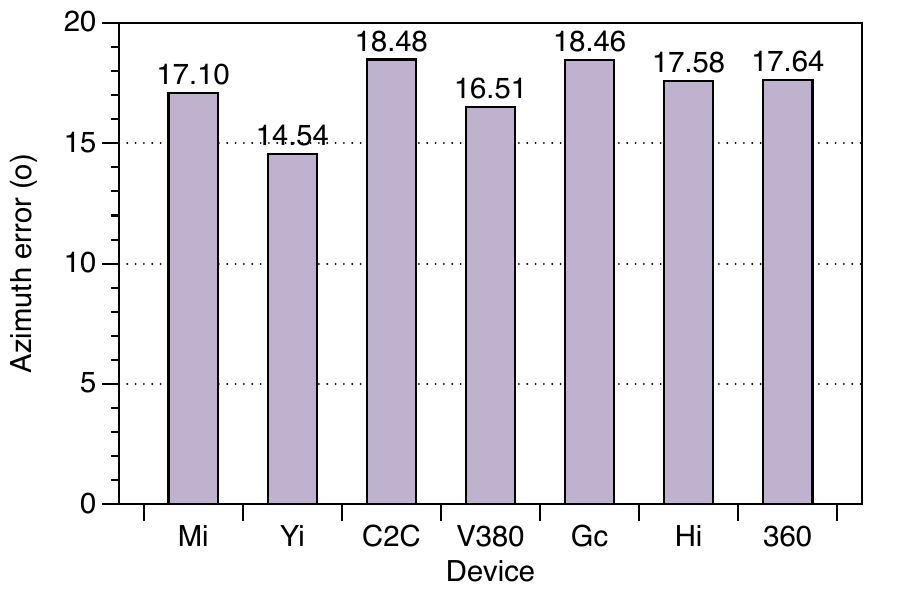}
  \vspace{-0.15in}
  \caption{Azimuth localization results across different device.}
  \label{fig:sendev}
  \vspace{-0.12in}
\end{figure}

\begin{figure}[]
  \centering
  \includegraphics[width=0.65\linewidth]{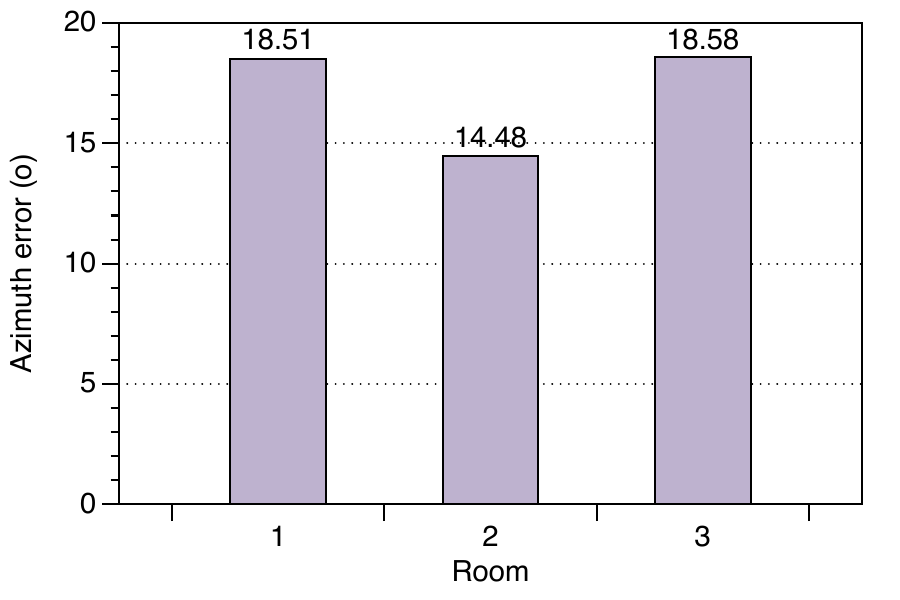}
  \vspace{-0.15in}
  \caption{Azimuth localization results across different room.}
  \label{fig:senroom}
  \vspace{-0.23in}
\end{figure}

\name's performance in localizing different devices is shown in Figure~\ref{fig:sendev}. Although the 360 camera's QoS data frames cannot be captured by common packet-sniffing tools, its CSI can still be monitored by the nexmon tool. Therefore, we also included the 360 camera in our localization experiments. As shown in Figure~\ref{fig:sendev}, \name maintains consistent localization performance across different types of cameras, demonstrating its robustness to device variations. Figure~\ref{fig:senroom} illustrates the differences in \name's azimuth localization performance across three rooms, further highlighting its robustness to environmental changes. This robustness stems from \name's localization algorithm, which is based on the analysis of wireless signal propagation paths. Consequently, unlike previous approaches~\cite{gu2024loccams}, it does not require pre-collected data or deep learning models that need extensive data to ensure robustness.

\section{Disscussion}
\label{sec:diss}

In this section, we discuss the limitations of \name, the potential risks, and possible improvements.

\noindent\textbf{Non-WiFi Cameras.} The fundamental principle behind \name's detection and localization of wireless cameras limits its applicability to live streaming spy cameras on WiFi networks. It does not extend to cameras that use local storage, cellular networks, or Ethernet. However, most recent crime cases have involved WiFi spy cameras~\cite{heo2022there} because they are easy to deploy and manage, and their prevalence is rapidly increasing in the commercial market. Therefore, \name is suitable for many scenarios. To expand the detection range, infrared or optical methods~\cite{sami2021lapd,yu2022heatdecam} would still be needed.

\noindent\textbf{MAC Address Randomization.} \name uses MAC addresses to identify devices and treats them as unique IDs throughout the localization process. Although some modern devices employ MAC address randomization~\cite{vanhoef2016mac} to enhance security, this does not affect \name's detection and localization capabilities. This is because devices, even with MAC address randomization, use a consistent MAC address for communication once a network connection is established.

\noindent\textbf{Non-VBR Devices.} When \name detects whether a camera is monitoring the current area, the device's traffic must be encoded using a Variable Bit Rate (VBR) algorithm. While this algorithm is used by the vast majority of wireless camera devices, if a camera is specifically designed to encode video/audio information at a constant bit rate (CBR), \name may only be able to roughly detect its presence using the OUI table. However, \name can still locate such devices through the proposed localization scheme.

\noindent\textbf{Evading \name.} \name aims to detect and locate wireless cameras deployed by typical attackers. However, we acknowledge that more powerful attackers may have ways to evade \name. Attackers could modify the behavior of hidden cameras by customizing hardware or altering firmware to change the packet size or arrival intervals, thus avoiding detection. For example, they could have the wireless camera transmit unencoded raw video, add padding traffic to ensure a stable data stream, use randomly varying resolutions, or inject noise to disrupt the traffic model. These methods could prevent \name from detecting whether a camera is monitoring the current area. However, such tactics require a high level of expertise from the attacker. The localization module, based on wireless signal propagation path analysis, can still function normally by using the device's MAC address and WiFi channel. Avoiding localization would require modifying the network card hardware to control the WiFi signal's transmission power, causing it to constantly change and disrupt the signal attenuation trend caused by user activity. This also requires attackers to have specialized knowledge, and modifying network card hardware is considerably challenging.

\begin{figure}[t]
  \centering
  \includegraphics[width=0.8\linewidth]{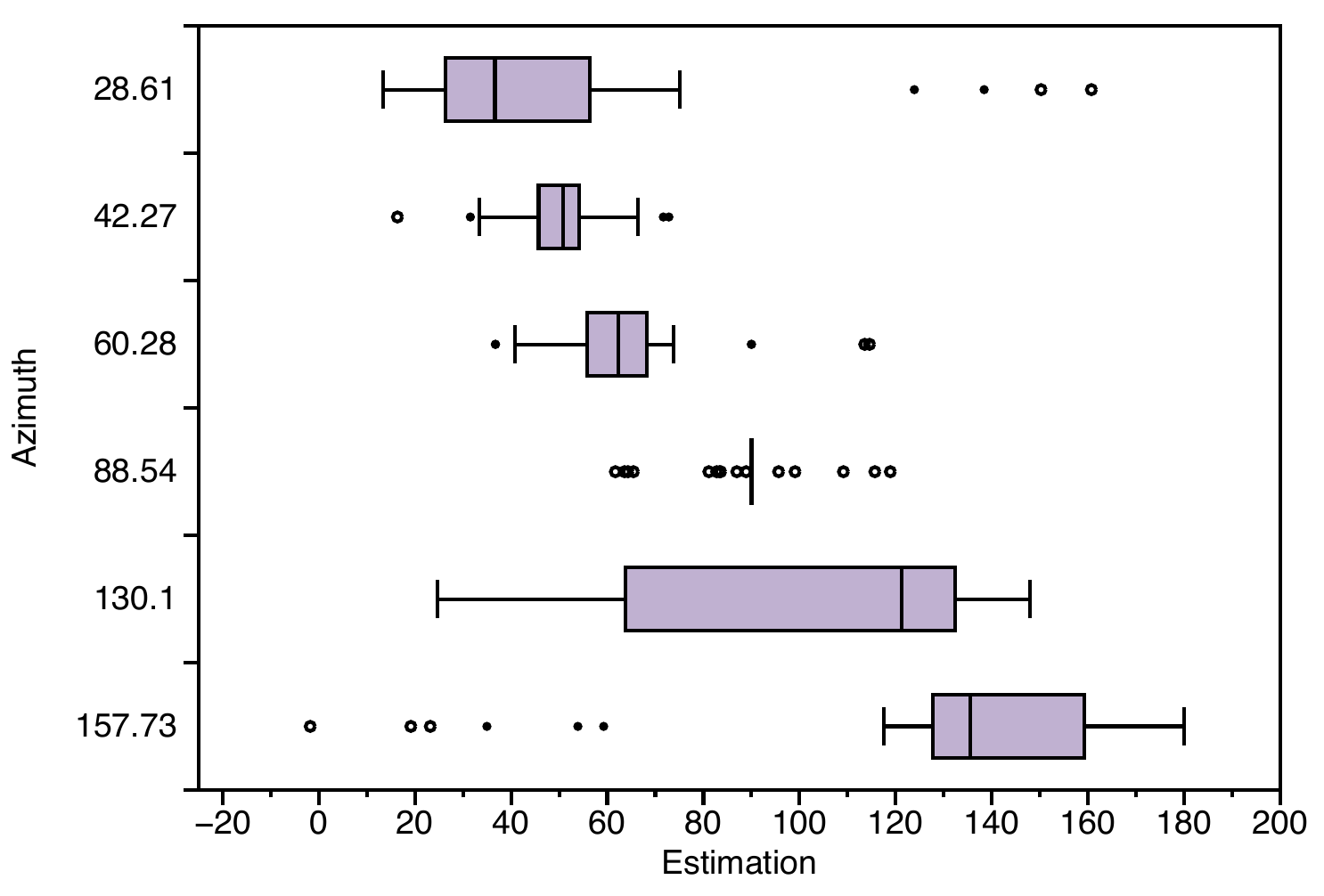}
  \vspace{-0.1in}
  \caption{Failures analysis.}
  \label{fig:falroom1}
  \vspace{-0.22in}
\end{figure}

\noindent\textbf{Failures Analysis.} To analyze the sources of error in \name's localization process, we present the localization results for Room 1 in Figure~\ref{fig:falroom1}. It can be observed that most errors originate from cameras with smaller azimuth angles and outliers. For cameras with smaller azimuth angles, the user's starting position pf Path2 is in front of the \name device, lacking the buffer distance similar to Path 1 (i.e., the user is already walking at a normal speed before entering the FFZ.) The differences in user acceleration and the limitation of leaving the FFZ right after starting to walk lead to a significant discrepancy between the actual and theoretical times for the user crossing the FFZ on Path 2. This discrepancy is difficult to correct with rules. Outliers mainly stem from the inaccuracy of the quadrant localization method. Due to the complexity of indoor multipath effects, the fluctuations in CSI 3 can be unstable, causing the algorithm to make incorrect judgments. This can result in a predicted angle that differs from the actual angle by $180 - \theta$. Therefore, in practical use, if the user does not find the hidden camera at the predicted location, they can also check the position at $180 - \theta$ to find camera.

\noindent\textbf{New Privacy Risks.} While \name can be used by users to detect wireless cameras, it could also be exploited by attackers to locate wireless devices in the user's environment through walls. Combined with traffic analysis and activity recognition methods based on WiFi CSI~\cite{chen2023cross,liu2023towards,yi2024bfmsense,li2024uwb,wang2024muki,zhang2023wifi}, this could create new privacy risks. This is a direction we intend to explore in our future research.

\noindent\textbf{Environment and User Effort Limitations.} \name can only localize wireless cameras within the 0-180$^o$ range. However, in real-world environments, it is relatively easy to find a location near a wall to place the \name device, and it can perform two rounds of positioning to achieve 360$^o$ localization. Therefore, this localization range remains useful in practical settings. Another limitation is that \name assumes users walk along two orthogonal straight paths at a constant speed, which can introduce errors in real-world scenarios. In actual environments, the layout of indoor furniture (such as floor stripes, walls, and furniture) can help guide users to maintain two straight walking paths. Additionally, users can easily control their walking speed within a certain range to minimize the biases. Our experiments have demonstrated the robustness of \name in real-world settings with normal user walking. \name can be easily extended to scenarios with multiple cameras. During the snooping camera detection phase, a single user walking can detect multiple cameras by clustering the MAC addresses of all sniffed packets. However, when capturing CSI, nexmon tool can only capture packets from one MAC address at a time. Therefore, to localize multiple cameras, the user must repeat the localization process for each camera, which increases the user's effort.

\noindent\textbf{Future Work for Improvement.} Next, we aim to further reduce user effort and eliminate localization errors caused by user activity. This will involve using low cost 3D-printed kits with metal obstructions as peripherals. By controlling the metal obstructions to rotate around the Raspberry Pi, we can perturb the CSI. Constructing a corresponding CSI-azimuth model will enable more precise localization with less user effort. Additionally, we plan to explore building indoor wireless device maps based on our localization technology. Combine this map with WiFi traffic and CSI will help us study new user behavior privacy risks and develop defensive measures.

\section{Conclusion}

In this paper, we propose \name, a framework for detecting and locating wireless hidden cameras based on wireless signal propagation path analysis, specifically focusing on diffraction attenuation. \name establishes a relationship between the signal attenuation caused by user activity and the location of the wireless camera. Compared to current methods, \name offers several advantages: it is cost-effective, requires no training, demands less activity space, and involves minimal user effort. However, \name still has some limitations. In future work, we aim to further reduce user effort and minimize localization errors through the use of peripherals. Additionally, we will analyze new privacy risks associated with \name and develop countermeasures.





%
\bibliographystyle{unsrt}
\bibliography{template}

\begin{thebibliography}{10}

\bibitem{ankit2024}
Ankit Gupta.
\newblock Wireless monitoring and surveillance market, by component, type,
  connectivity, end-user - forecast till 2030.
\newblock
  \url{https://www.marketresearchfuture.com/reports/wireless-monitoring-surveillance-market-975},
  2024.

\bibitem{ye2013wireless}
Yun Ye, Song Ci, Aggelos~K Katsaggelos, Yanwei Liu, and Yi~Qian.
\newblock Wireless video surveillance: A survey.
\newblock {\em IEEE Access}, 1:646--660, 2013.

\bibitem{mare2020smart}
Shrirang Mare, Franziska Roesner, and Tadayoshi Kohno.
\newblock Smart devices in airbnbs: Considering privacy and security for both
  guests and hosts.
\newblock {\em Proceedings on Privacy Enhancing Technologies}, 2020.

\bibitem{David2023}
David Janssen.
\newblock Many airbnbs have cameras installed, especially in the us, canada and
  singapore.
\newblock \url{}, 2023.

\bibitem{Jim2019}
Jim Dalrymple.
\newblock More than 1 in 10 airbnb guests have found hidden cameras: Survey.
\newblock
  \url{https://www.inman.com/2019/06/07/morethan-1-in-10-airbnb-guest-have-found-cameras-in-rentals-survey/},
  2019.

\bibitem{delaware}
The security camera laws in delaware.
\newblock
  \url{https://www.cambasket.com/the-security-camera-laws-in-delaware/}.

\bibitem{sami2021lapd}
Sriram Sami, Sean Rui~Xiang Tan, Bangjie Sun, and Jun Han.
\newblock Lapd: Hidden spy camera detection using smartphone time-of-flight
  sensors.
\newblock In {\em Proceedings of the 19th ACM Conference on Embedded Networked
  Sensor Systems}, pages 288--301, 2021.

\bibitem{yu2022heatdecam}
Zhiyuan Yu, Zhuohang Li, Yuanhaur Chang, Skylar Fong, Jian Liu, and Ning Zhang.
\newblock Heatdecam: detecting hidden spy cameras via thermal emissions.
\newblock In {\em Proceedings of the 2022 ACM SIGSAC Conference on Computer and
  Communications Security}, pages 3107--3120, 2022.

\bibitem{sharma2022lumos}
Rahul~Anand Sharma, Elahe Soltanaghaei, Anthony Rowe, and Vyas Sekar.
\newblock Lumos: Identifying and localizing diverse hidden $\{$IoT$\}$ devices
  in an unfamiliar environment.
\newblock In {\em 31st USENIX Security Symposium (USENIX Security 22)}, pages
  1095--1112, 2022.

\bibitem{singh2021always}
Akash~Deep Singh, Luis Garcia, Joseph Noor, and Mani Srivastava.
\newblock I always feel like somebody's sensing me! a framework to detect,
  identify, and localize clandestine wireless sensors.
\newblock In {\em 30th USENIX Security Symposium (USENIX Security 21)}, pages
  1829--1846, 2021.

\bibitem{he2021motioncompass}
Yan He, Qiuye He, Song Fang, and Yao Liu.
\newblock Motioncompass: pinpointing wireless camera via motion-activated
  traffic.
\newblock In {\em Proceedings of the 19th Annual International Conference on
  Mobile Systems, Applications, and Services}, pages 215--227, 2021.

\bibitem{heo2022there}
Jeongyoon Heo, Sangwon Gil, Youngman Jung, Jinmok Kim, Donguk Kim, Woojin Park,
  Yongdae Kim, Kang~G Shin, and Choong-Hoon Lee.
\newblock Are there wireless hidden cameras spying on me?
\newblock In {\em Proceedings of the 38th Annual Computer Security Applications
  Conference}, pages 714--726, 2022.

\bibitem{gu2024loccams}
Yangyang Gu, Jing Chen, Cong Wu, Kun He, Ziming Zhao, and Ruiying Du.
\newblock Loccams: An efficient and robust approach for detecting and
  localizing hidden wireless cameras via commodity devices.
\newblock {\em Proceedings of the ACM on Interactive, Mobile, Wearable and
  Ubiquitous Technologies}, 7(4):1--24, 2024.

\bibitem{cheng2018dewicam}
Yushi Cheng, Xiaoyu Ji, Tianyang Lu, and Wenyuan Xu.
\newblock Dewicam: Detecting hidden wireless cameras via smartphones.
\newblock In {\em Proceedings of the 2018 on Asia Conference on Computer and
  Communications Security}, pages 1--13, 2018.

\bibitem{wu2019you}
Kevin Wu and Brent Lagesse.
\newblock Do you see what i see? detecting hidden streaming cameras through
  similarity of simultaneous observation.
\newblock In {\em 2019 IEEE International Conference on Pervasive Computing and
  Communications (PerCom}, pages 1--10. IEEE, 2019.

\bibitem{ji2018user}
Xiaoyu Ji, Yushi Cheng, Wenyuan Xu, and Xinyan Zhou.
\newblock User presence inference via encrypted traffic of wireless camera in
  smart homes.
\newblock {\em Security and Communication Networks}, 2018(1):3980371, 2018.

\bibitem{cheng2019detecting}
Yushi Cheng, Xiaoyu Ji, Tianyang Lu, and Wenyuan Xu.
\newblock On detecting hidden wireless cameras: A traffic pattern-based
  approach.
\newblock {\em IEEE Transactions on Mobile Computing}, 19(4):907--921, 2019.

\bibitem{salman2022csi}
Muhammad Salman, Nguyen Dao, Uichin Lee, and Youngtae Noh.
\newblock Csi: Despy: enabling effortless spy camera detection via passive
  sensing of user activities and bitrate variations.
\newblock {\em Proceedings of the ACM on Interactive, Mobile, Wearable and
  Ubiquitous Technologies}, 6(2):1--27, 2022.

\bibitem{dao2021deepdespy}
Dinhnguyen Dao, Muhammad Salman, and Youngtae Noh.
\newblock Deepdespy: a deep learning-based wireless spy camera detection
  system.
\newblock {\em IEEE Access}, 9:145486--145497, 2021.

\bibitem{jakobispy2023}
Jakobi Teknik.
\newblock Spy hidden camera detector.
\newblock \url{https://apps.apple.com/us/app/spy-hidden-cameradetector/id9259
  67783?mt=8}, 2023.

\bibitem{llchidden2023}
LLC LSC.
\newblock Hidden camera detector.
\newblock
  \url{https://apps.apple.com/us/app/hidden-camera-detector/id532882360}, 2023.

\bibitem{liu2023camradar}
Ziwei Liu, Feng Lin, Chao Wang, Yijie Shen, Zhongjie Ba, Li~Lu, Wenyao Xu, and
  Kui Ren.
\newblock Camradar: hidden camera detection leveraging amplitude-modulated
  sensor images embedded in electromagnetic emanations.
\newblock {\em Proceedings of the ACM on Interactive, Mobile, Wearable and
  Ubiquitous Technologies}, 6(4):1--25, 2023.

\bibitem{zuniga2022see}
Agustin Zuniga, Naser~Hossein Motlagh, Mohammad~A Hoque, Sasu Tarkoma, Huber
  Flores, and Petteri Nurmi.
\newblock See no evil: Discovering covert surveillance devices using thermal
  imaging.
\newblock {\em IEEE Pervasive Computing}, 21(4):33--42, 2022.

\bibitem{ma2023lenser}
Yongqiang Ma, Xiangyang Luo, Ruixiang Li, Shaoyong Du, and Wenyan Liu.
\newblock Lenser: A channel state information based indoor localization scheme
  for malicious devices.
\newblock In {\em 2023 IEEE 20th International Conference on Mobile Ad Hoc and
  Smart Systems (MASS)}, pages 461--470. IEEE, 2023.

\bibitem{baker2003mathematical}
Bevan~B Baker and Edward~Thomas Copson.
\newblock {\em The mathematical theory of Huygens' principle}, volume 329.
\newblock American Mathematical Soc., 2003.

\bibitem{goldsmith2005wireless}
Andrea Goldsmith.
\newblock {\em Wireless communications}.
\newblock Cambridge university press, 2005.

\bibitem{sathyamoorthy2014wireless}
Dinesh Sathyamoorthy, Mohd Jalis~Md Jelas, and Shalini Shafii.
\newblock Wireless spy devices: A review of technologies and detection methods.
\newblock {\em Editorial Board}, 7:130, 2014.

\bibitem{valeros2017spy}
Veronica Valeros and Sebastian Garcia.
\newblock Spy vs. spy: A modern study of microphone bugs operation and
  detection.
\newblock {\em Chaos Computer Club eV}, 2017.

\bibitem{liu2018detecting}
Tian Liu, Ziyu Liu, Jun Huang, Rui Tan, and Zhen Tan.
\newblock Detecting wireless spy cameras via stimulating and probing.
\newblock In {\em Proceedings of the 16th Annual International Conference on
  Mobile Systems, Applications, and Services}, pages 243--255, 2018.

\bibitem{miettinen2017iot}
Markus Miettinen, Samuel Marchal, Ibbad Hafeez, N~Asokan, Ahmad-Reza Sadeghi,
  and Sasu Tarkoma.
\newblock Iot sentinel: Automated device-type identification for security
  enforcement in iot.
\newblock In {\em 2017 IEEE 37th international conference on distributed
  computing systems (ICDCS)}, pages 2177--2184. IEEE, 2017.

\bibitem{huang2023phyfinatt}
Jinyang Huang, Bin Liu, Chenglin Miao, Xiang Zhang, Jiancun Liu, Lu~Su, Zhi
  Liu, and Yu~Gu.
\newblock Phyfinatt: An undetectable attack framework against phy layer
  fingerprint-based wifi authentication.
\newblock {\em IEEE Transactions on Mobile Computing}, 2023.

\bibitem{zhang2023wital}
Xiang Zhang, Yu~Gu, Huan Yan, Yantong Wang, Mianxiong Dong, Kaoru Ota, Fuji
  Ren, and Yusheng Ji.
\newblock Wital: A cots wifi devices based vital signs monitoring system using
  nlos sensing model.
\newblock {\em IEEE Transactions on Human-Machine Systems}, 53(3):629--641,
  2023.

\bibitem{zhang2024wiopen}
Xiang Zhang, Jingyang Huang, Huan Yan, Peng Zhao, Guohang Zhuang, Zhi Liu, and
  Bin Liu.
\newblock Wiopen: A robust wi-fi-based open-set gesture recognition framework.
\newblock {\em arXiv preprint arXiv:2402.00822}, 2024.

\bibitem{gu2023wife}
Yu~Gu, Xiang Zhang, Huan Yan, Jingyang Huang, Zhi Liu, Mianxiong Dong, and Fuji
  Ren.
\newblock Wife: Wifi and vision based unobtrusive emotion recognition via
  gesture and facial expression.
\newblock {\em IEEE Transactions on Affective Computing}, 2023.

\bibitem{rappaport2024wireless}
Theodore~S Rappaport.
\newblock {\em Wireless communications: principles and practice}.
\newblock Cambridge University Press, 2024.

\bibitem{zhang2019towards}
Fusang Zhang, Kai Niu, Jie Xiong, Beihong Jin, Tao Gu, Yuhang Jiang, and Daqing
  Zhang.
\newblock Towards a diffraction-based sensing approach on human activity
  recognition.
\newblock {\em Proceedings of the ACM on Interactive, Mobile, Wearable and
  Ubiquitous Technologies}, 3(1):1--25, 2019.

\bibitem{yao2024wiprofile}
Zhiyun Yao, Xuanzhi Wang, Kai Niu, Rong Zheng, Junzhe Wang, and Daqing Zhang.
\newblock Wiprofile: Unlocking diffraction effects for sub-centimeter target
  profiling using commodity wifi devices.
\newblock In {\em Proceedings of the 30th Annual International Conference on
  Mobile Computing and Networking}, pages 185--199, 2024.

\bibitem{wang2024understanding}
Xuanzhi Wang, Anlan Yu, Kai Niu, Weiyan Shi, Junzhe Wang, Zhiyun Yao, Rahul~C
  Shah, Hong Lu, and Daqing Zhang.
\newblock Understanding the diffraction model in static multipath-rich
  environments for wifi sensing system design.
\newblock {\em IEEE Transactions on Mobile Computing}, 2024.

\bibitem{halperin2011tool}
Daniel Halperin, Wenjun Hu, Anmol Sheth, and David Wetherall.
\newblock Tool release: Gathering 802.11 n traces with channel state
  information.
\newblock {\em ACM SIGCOMM computer communication review}, 41(1):53--53, 2011.

\bibitem{lireshaping}
Rui Li, Yu~Duan, Rui Du, Fangxin Xu, Hangbin Zhao, Yang Sun, Yiyan Zhang,
  Daiyang Zhang, Yiming Liu, Zhiping Jiang, et~al.
\newblock Reshaping wi-fi isac with high-coherence hardware capabilities.

\bibitem{gringoli2019free}
Francesco Gringoli, Matthias Schulz, Jakob Link, and Matthias Hollick.
\newblock Free your csi: A channel state information extraction platform for
  modern wi-fi chipsets.
\newblock In {\em Proceedings of the 13th International Workshop on Wireless
  Network Testbeds, Experimental Evaluation \& Characterization}, pages 21--28,
  2019.

\bibitem{nexmon:project}
Matthias Schulz, Daniel Wegemer, and Matthias Hollick.
\newblock Nexmon: The c-based firmware patching framework.
\newblock \url{https://nexmon.org}, 2017.

\bibitem{ma2019wifi}
Yongsen Ma, Gang Zhou, and Shuangquan Wang.
\newblock Wifi sensing with channel state information: A survey.
\newblock {\em ACM Computing Surveys (CSUR)}, 52(3):1--36, 2019.

\bibitem{gu2022wigrunt}
Yu~Gu, Xiang Zhang, Yantong Wang, Meng Wang, Huan Yan, Yusheng Ji, Zhi Liu,
  Jianhua Li, and Mianxiong Dong.
\newblock Wigrunt: Wifi-enabled gesture recognition using dual-attention
  network.
\newblock {\em IEEE Transactions on Human-Machine Systems}, 52(4):736--746,
  2022.

\bibitem{wampler2015information}
Christopher Wampler, Selcuk Uluagac, and Raheem Beyah.
\newblock Information leakage in encrypted ip video traffic.
\newblock In {\em 2015 IEEE Global Communications Conference (GLOBECOM)}, pages
  1--7. IEEE, 2015.

\bibitem{nassi2019drones}
Ben Nassi, Raz Ben-Netanel, Adi Shamir, and Yuval Elovici.
\newblock Drones' cryptanalysis-smashing cryptography with a flicker.
\newblock In {\em 2019 IEEE Symposium on Security and Privacy (SP)}, pages
  1397--1414. IEEE, 2019.

\bibitem{Fussell2019}
S.~Fussell.
\newblock Airbnb has a hidden-camera problem.
\newblock \url{https://www.theatlantic.com/
  technology/archive/2019/03/what-happens-when-youfind-cameras-your-airbnb/585007/},
  2024.

\bibitem{Jeong2019}
S.~Jeong and J.~Griffiths.
\newblock Hundreds of south korean motel guests were secretly filmed and
  live-streamed online.
\newblock
  \url{https://www.cnn.com/2019/03/20/asia/southkorea-hotel-spy-cam-intl/index.html},
  2019.

\bibitem{10587029}
Jinyang Huang, Jia-Xuan Bai, Xiang Zhang, Zhi Liu, Yuanhao Feng, Jianchun Liu,
  Xiao Sun, Mianxiong Dong, and Meng Li.
\newblock Keystrokesniffer: An off-the-shelf smartphone can eavesdrop on your
  privacy from anywhere.
\newblock {\em IEEE Transactions on Information Forensics and Security}, pages
  1--1, 2024.

\bibitem{ortiz2019devicemien}
Jorge Ortiz, Catherine Crawford, and Franck Le.
\newblock Devicemien: network device behavior modeling for identifying unknown
  iot devices.
\newblock In {\em Proceedings of the International Conference on Internet of
  Things Design and Implementation}, pages 106--117, 2019.

\bibitem{sivanathan2018classifying}
Arunan Sivanathan, Hassan~Habibi Gharakheili, Franco Loi, Adam Radford, Chamith
  Wijenayake, Arun Vishwanath, and Vijay Sivaraman.
\newblock Classifying iot devices in smart environments using network traffic
  characteristics.
\newblock {\em IEEE Transactions on Mobile Computing}, 18(8):1745--1759, 2018.

\bibitem{meta2024}
Metageek.
\newblock The basics: Understanding rssi.
\newblock
  \url{https://www.metageek.com/training/resources/understanding-rssi/}, 2019.

\bibitem{li2022packet}
Jianfeng Li, Shuohan Wu, Hao Zhou, Xiapu Luo, Ting Wang, Yangyang Liu, and
  Xiaobo Ma.
\newblock Packet-level open-world app fingerprinting on wireless traffic.
\newblock In {\em The 2022 Network and Distributed System Security Symposium
  (NDSS'22)}, 2022.

\bibitem{gast2005802}
Matthew Gast.
\newblock {\em 802.11 wireless networks: the definitive guide}.
\newblock O'Reilly Media, Inc., 2005.

\bibitem{9363693}
Ieee standard for information technology--telecommunications and information
  exchange between systems - local and metropolitan area networks--specific
  requirements - part 11: Wireless lan medium access control (mac) and physical
  layer (phy) specifications.
\newblock {\em IEEE Std 802.11-2020 (Revision of IEEE Std 802.11-2016)}, pages
  1--4379, 2021.

\bibitem{van2008traffic}
Geert Van~der Auwera, Prasanth~T David, and Martin Reisslein.
\newblock Traffic characteristics of h. 264/avc variable bit rate video.
\newblock {\em IEEE Communications Magazine}, 46(11):164--174, 2008.

\bibitem{pan2016fast}
Zhaoqing Pan, Jianjun Lei, Yun Zhang, Xingming Sun, and Sam Kwong.
\newblock Fast motion estimation based on content property for low-complexity
  h. 265/hevc encoder.
\newblock {\em IEEE Transactions on Broadcasting}, 62(3):675--684, 2016.

\bibitem{zhang2018}
Fusang Zhang, Daqing Zhang, Jie Xiong, Hao Wang, Kai Niu, Beihong Jin, and
  Yuxiang Wang.
\newblock From fresnel diffraction model to fine-grained human respiration
  sensing with commodity wi-fi devices.
\newblock {\em Proc. ACM Interact. Mob. Wearable Ubiquitous Technol.}, 2(1),
  mar 2018.

\bibitem{vanhoef2016mac}
Mathy Vanhoef, C{\'e}lestin Matte, Mathieu Cunche, Leonardo~S Cardoso, and
  Frank Piessens.
\newblock Why mac address randomization is not enough: An analysis of wi-fi
  network discovery mechanisms.
\newblock In {\em Proceedings of the 11th ACM on Asia conference on computer
  and communications security}, pages 413--424, 2016.

\bibitem{chen2023cross}
Chen Chen, Gang Zhou, and Youfang Lin.
\newblock Cross-domain wifi sensing with channel state information: A survey.
\newblock {\em ACM Computing Surveys}, 55(11):1--37, 2023.

\bibitem{liu2023towards}
Jinyi Liu, Wenwei Li, Tao Gu, Ruiyang Gao, Bin Chen, Fusang Zhang, Dan Wu, and
  Daqing Zhang.
\newblock Towards a dynamic fresnel zone model to wifi-based human activity
  recognition.
\newblock {\em Proceedings of the ACM on Interactive, Mobile, Wearable and
  Ubiquitous Technologies}, 7(2):1--24, 2023.

\bibitem{yi2024bfmsense}
Enze Yi, Dan Wu, Jie Xiong, Fusang Zhang, Kai Niu, Wenwei Li, and Daqing Zhang.
\newblock $\{$BFMSense$\}$:$\{$WiFi$\}$ sensing using beamforming feedback
  matrix.
\newblock In {\em 21st USENIX Symposium on Networked Systems Design and
  Implementation (NSDI 24)}, pages 1697--1712, 2024.

\bibitem{li2024uwb}
Xin Li, Hongbo Wang, Zhe Chen, Zhiping Jiang, and Jun Luo.
\newblock Uwb-fi: Pushing wi-fi towards ultra-wideband for fine-granularity
  sensing.
\newblock In {\em Proceedings of the 22nd Annual International Conference on
  Mobile Systems, Applications and Services}, pages 42--55, 2024.

\bibitem{wang2024muki}
Hongbo Wang, Jingyang Hu, Tianyue Zheng, Jingzhi Hu, Zhe Chen, Hongbo Jiang,
  Yuanjin Zheng, and Jun Luo.
\newblock Muki-fi: Multi-person keystroke inference with bfi-enabled wi-fi
  sensing.
\newblock {\em IEEE Transactions on Mobile Computing}, 2024.

\bibitem{zhang2023wifi}
Daqing Zhang, Kai Niu, Jie Xiong, Fusang Zhang, and Xuanzhi Wang.
\newblock Wifi/4g/5g based wireless sensing: Theories, applications and future
  directions.
\newblock In {\em Integrated Sensing and Communications}, pages 387--417.
  Springer, 2023.

\end{thebibliography}

\end{document}